\documentclass[sn-mathphys-num,iicol]{sn-jnl}

\usepackage{float}
\usepackage{soul}
\usepackage{xcolor}
\usepackage{physics}
\usepackage{relsize}
\usepackage{makecell}

\usepackage{amsmath}
\usepackage{amsfonts}
\usepackage{amssymb}
\usepackage{mathrsfs}

\usepackage{mathtools}
\usepackage{caption}
\usepackage{subcaption}

\newcommand{\figref}{Figure }
\newcommand{\vout}[1]{\mathop{\mathbb{#1}}} 
\newcommand{\vcal}[1]{\mathcal{\mathbb{#1}}} 
\newcommand{\FET}{FE\textsuperscript{2}}
\newcommand{\mumfim}{MuMFiM}

\newcommand{\rev}[1]{#1}
\newcommand{\revtwo}[1]{#1}

\newcommand{\seng}{\mathcal{W}}
\newcommand{\sstr}{\vb{\sigma}}
\newcommand{\srcg}{\vb{C}}
\newcommand{\sstiff}{\vout{D}}
\newcommand{\slossk}[1]{\mathcal{L}^{#1}}

\begin{document}

\author[1]{\fnm{Nishan} \sur{Parvez}}\email{parvem@rpi.edu}
\author*[1]{\fnm{Jacob} \sur{Merson}}\email{mersoj2@rpi.edu}

\affil[1]{\orgdiv{Department of Mechanical, Aerospace, and Nuclear Engineering}, \orgname{Rensselaer Polytechnic Institute}, \orgaddress{\street{110 8th St.} \city{Troy}, \state{New York}, \postcode{12180}, \country{USA}}}

\title[A Physics Preserving Neural Network Based Approach for Constitutive Modeling of Isotropic Fibrous Materials]{A Physics Preserving Neural Network Based Approach for Constitutive Modeling of Isotropic Fibrous Materials}


\abstract{
We develop a new neural network based material model for discrete fibrous materials that strictly enforces constitutive constraints such as polyconvexity, frame-indifference, and the symmetry of the stress and material stiffness. Additionally, we show that the accuracy of the stress and material stiffness predictions is significantly improved for this neural network by using a Sobolev minimization strategy that includes derivative terms. We obtain a normalized mean square error of 0.15\% for the strain energy density, 0.815\% averaged across the components of the stress, and 5.4\% averaged across the components of the stiffness tensor. This machine-learned constitutive model was deployed in a finite element simulation of a facet capsular ligament. The displacement fields and stress-strain curves were compared to a multiscale simulation that required running on a GPU-based supercomputer. The new approach maintained upward of 85\% accuracy in stress up to 70\% strain while reducing the computation cost by orders of magnitude.}

\date{Version \versionno, \today}

\maketitle

\section{Introduction}
Biological materials are among the most complex materials to model due to their microstructural heterogeneity, innate length-scale spanning orders of magnitude, and affinity for large deformation. Alongside traditional analytical formulation of hyperelastic material models, these complexities have led to the emergence of various multiscale methodologies that enable a greater understanding of the interplay between scales. They are gaining particular traction in modeling biological tissues with a network-like microstructure made of collagen fibers \cite{mersonNewOpensourceFramework2024, mahutgaNonaffineFiberNetwork2023, laiMultiscaleApproachModeling2013}.

An upscaling multiscale approach is typical for modeling biological tissues due to a strong scale separation between the macroscale (e.g., model) and the subscale (e.g., fiber) length scales. In these methods, the subscale material behavior is homogenized to provide the constitutive properties for the macroscale analysis. This homogenization can either be done concurrently with the macroscale solution or a priori.

In a priori homogenization, numerical procedures such as the finite element method are typically used to perform a standard set of computational experiments (e.g., uniaxial displacement, biaxial displacement, pure shear, etc.) on a discrete material model to calibrate an analytical constitutive model, such as the neo-Hookean, HGO, or Fung model \cite{holzapfelNewConstitutiveFramework2000, mellyReviewMaterialModels2021}. These constitutive models generally contain a small number of free parameters that need to be calibrated. As a result, this approach is computationally inexpensive because the needed set of numerical experiments is kept to a minimum. 

Another option is to use an upscaling multiscale method where the microstructure is homogenized concurrently with the macroscale solution. One example of this scheme is the \FET\ method developed by \cite{feyelMultiscaleFE2Elastoviscoplastic1999} and others. In this procedure, the representative microstructures are simulated concurrently using the deformation gradient informed by the macroscale problem (i.e., the deformation gradient experienced by the finite elements of the macroscale model). This mitigates the need for an analytical constitutive response function and can handle a broader range of emergent behaviors. The microstructure may be a continuum or a discrete model, with the latter being the case for soft fibrous materials. 

Both a priori and concurrent hierarchical methods require strong scale separation and homogenization over a large enough domain that has representative material properties (i.e., the representative volume element (RVE)). The homogenization literature typically describes this as the size where the constitutive response remains unchanged under equivalent loading conditions (e.g., the Voigt and Reuss bounds converge). In 3D Voronoi networks, a common proxy for collagen networks, the RVE size is $\gg 40$ times the mean fiber length \cite{mersonSizeEffectsRandom2020}. However, microscale simulations of this size are not tractable in the context of concurrent upscaling multiscale analysis without the use of supercomputers and specialized solution procedures such as \mumfim\ described in \cite{mersonUsingHierarchicalParallelism2023}.

Due to this cost, the most common approach is to use a priori homogenization methods. The analytical material models for fibrous structures are based on the affine assumption that the fiber's end-to-end vector will deform according to a global deformation gradient. However, many authors have shown that the non-affine behavior of stochastic networks is critical to the overall response \cite{picuNetworkMaterialsStructure2022, hatami-marbiniEffectFiberOrientation2009,broederszMolecularMotorsStiffen2011,lakeEvaluationAffineFiber2012,leePresenceAffineFibril2015}. Consequently, traditional analytical constitutive response functions cannot capture behaviors that stem from micromechanical details such as dynamic realignment of fibers and non-locality, which are prominent in biological and polymeric materials and lead to emergent properties. One such observed emergent behavior is that networks exhibit Poisson's ratios over the thermodynamic limit for homogeneous elastic materials \cite{picuPoissonContractionFiber2018}. This behavior is modified when fibers are embedded into a matrix and is a strong function of the ratio of the fiber and matrix stiffnesses \cite{kakaletsisMechanicsEmbeddedFiber2023,deyEvaluationParallelCoupling2024}. Fortunately, data-driven techniques such as neural network-based models present an alternative pathway to develop material models without these limiting assumptions, which is the topic of the current study. 

Over the past decade, machine learning models have become an increasingly important part of the engineering design process due to promising accuracy in predicting complex deformation fields and in constitutive modeling of nonlinear materials \cite{thakolkaranNNEUCLIDDeeplearningHyperelasticity2022, sacksNeuralNetworkApproaches2022, kakaletsisCanMachineLearning2023, guptaAcceleratedMultiscaleMechanics2023, pengMultiscaleModelingMeets2021, lengPredictingMechanicalProperties2021}. Mathematically, a neural network is generally an involved composite function that contains controlled nonlinearity and a large number of free parameters that are calibrated to model a plethora of phenomena. The mathematical formulation is generally discussed in terms of ``neural network architecture'' and can be chosen so that the underlying functional form is convex (e.g., Input Convex Neural Network (ICNN) \cite{amosInputConvexNeural2017}, which is particularly conducive to continuum mechanics based material model formulation). Once trained, these surrogate models have computational efficiency comparable to traditional hyperelastic models while being more capable of handling nuances of complex microstructural deformations. In this paper, we adopt this framework to model fibrous network material using an extensive dataset obtained using \mumfim. \rev{Our main goal is to accelerate numerical simulations of multiscale models rather than to fit experimentally obtained data as is typical in the literature, for example, \cite{lengPredictingMechanicalProperties2021,thakolkaranNNEUCLIDDeeplearningHyperelasticity2022,tacDatadrivenModelingMechanical2022,linkaNewFamilyConstitutive2023}. This choice affords a different set of modeling constraints where we seek to obtain the highest quality downstream finite element solution even when it requires data that would be difficult or impossible to obtain experimentally---like full-field stress and material stiffness measurements.}

\rev{
The desire for polyconvex energy functions has been discussed at length in the constitutive modeling literature as it guarantees the ellipticity of the finite element problem and the existence of a minimizer of the elastic free energy and is consistent with objectivity unlike strict convexity \cite{hartmannPolyconvexityGeneralizedPolynomialtype2003,simoComputationalInelasticity1998}. There are two main ways in which this constraint is included in neural networks. First, the so-called ``Physics-Informed'' approach where convexity is added as a constraint to the loss function. This is the approach that is taken in \cite{tacDatadrivenModelingMechanical2022,lengPredictingMechanicalProperties2021}.
}

\rev{
An alternative approach is to strictly enforce polyconvexity through the network architecture. Simply put, the network architecture must guarantee convexity of the deformation gradient (\(\vb{F}\)), cofactor of \(\vb{F}\), and the determinant of \(\vb{F}\) or accounting for frame indifference convexity with respect to the components of the right Cauchy-Green deformation tensor (\(\vb{C}\), \(\text{cof}\, \vb{C}\), and \(\det \vb{C}\) ) \cite{truesdellNonLinearFieldTheories2004}. This approach is taken in \cite{kleinPolyconvexAnisotropicHyperelasticity2022,chenPolyconvexNeuralNetworks2022a,linkaNewFamilyConstitutive2023}. We also use this approach here, preferring strongly enforced constraints to weakly enforced constraints.
}

The primary contribution of this work is a comprehensive numerical material model describing the constitutive response of fibrous material without limiting assumptions such as affine deformation or incompressibility. The model maps the deformation gradient to the strain energy density function from which other relevant quantities, such as stress and stiffness tensor, are obtained. The particular choice of the neural network ensures the resulting model is compatible with the continuum mechanics framework. We further augment the material model's accuracy through the so-called Sobolev training protocol that considers stress and/or stiffness during ICNN training to obtain noticeably better predictions than ICNN trained using strain energy information only.

The overall computational framework we adopt here targets the finite element-based analysis of problems of practical interest. As a demonstration, we replace the computationally expensive microscale simulation of \mumfim\ package with the trained model to analyze the deformation of facet capsule ligament (FCL) and compare the result to that obtained using a concurrently homogenized approach that requires modern GPU-based supercomputers. The new paradigm reduces the computational costs by orders of magnitude while maintaining upward of 85\% accuracy in stress response up to 70\% uniaxial strain. Thus, we expect that the present work will enable researchers without access to specialized resources to investigate engineering problems involving fibrous materials with ease.

\section{Methods}
The numerical constitutive model developed here is purposefully designed to work with numerical procedures such as finite element methods. Thus, we describe the model's merits and place in a typical workflow involving fibrous materials by summarizing the relevant components of the overall analysis trajectory.

\begin{figure}
    \centering
    \includegraphics[width = \linewidth]{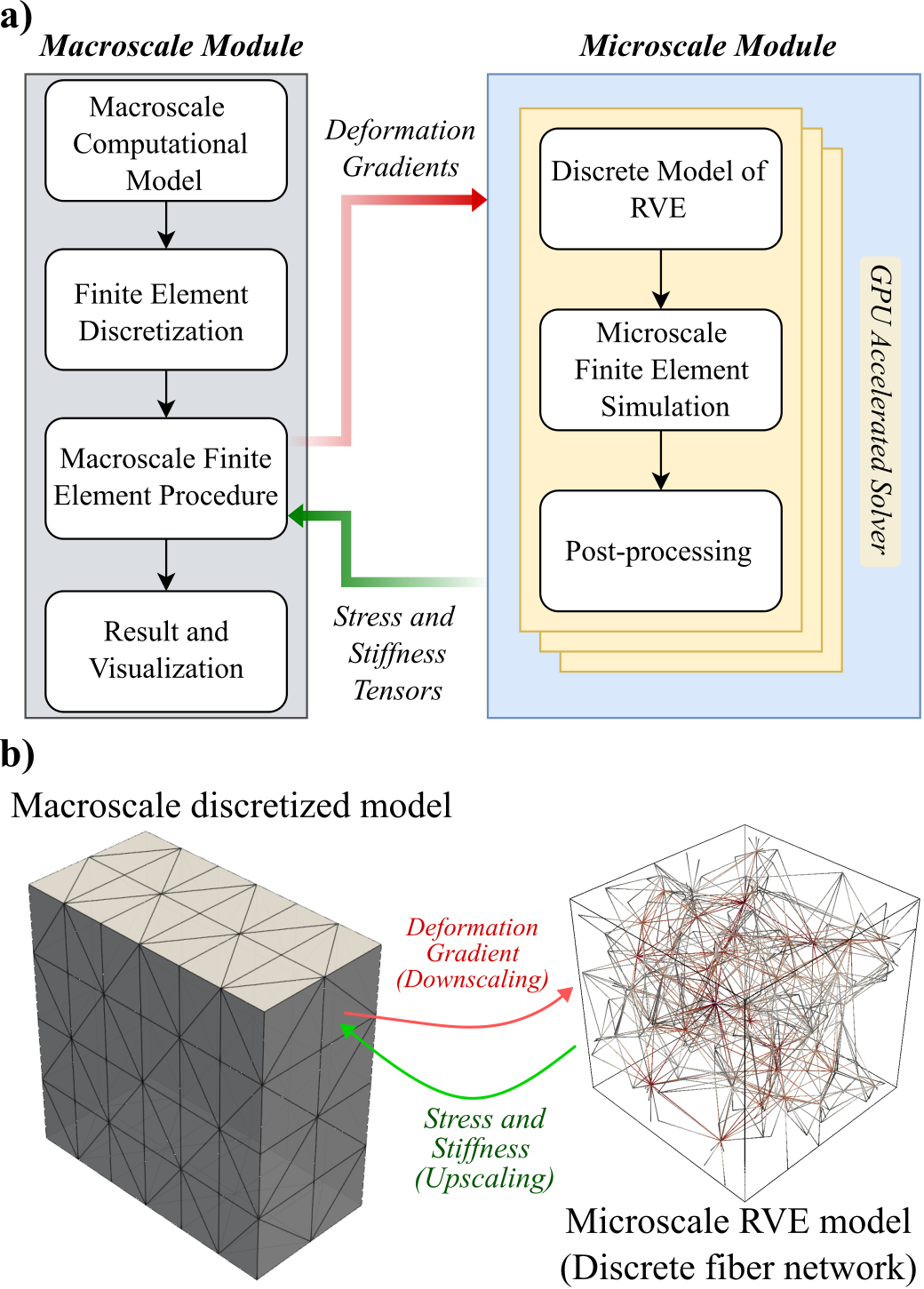}
    \caption{(a) Schematic of a typical workflow of \mumfim\ multiscale procedure that involves two finite element solution modules that handle the macro and microscale part of the problem, respectively. The deformation gradient, as informed by the macroscale module, is used to simulate RVEs to estimate the material response. For fibrous materials, the RVE is a discrete fiber network model (e.g., 3D Delaunay network shown in (b)). The machine learning model presented in this study is intended to be an accurate model of the RVE and aims to replace the `microscale` module in panel (a).}
    \label{fig:rve_schematic}
\end{figure}

For macroscale problems of practical interest, a multiscale method such as \FET \cite{feyelMultiscaleFE2Elastoviscoplastic1999} is generally necessary due to the microstructural complexities of the fibrous material. A simplified schematic of this approach is shown in \figref \ref{fig:rve_schematic}a. In this workflow, two finite-element solution modules are utilized. The first module analyzes the discretized macroscale problem, hereafter denoted as the macroscale module. The deformation gradient experienced by a macroscale element is applied to a representative volume element (RVE) by a second finite element solver, hereafter referred to as a microscale module. The output of this module is used to compute material response in the form of stress and stiffness tensor and passed back to the macroscale solver. The process is repeated for all finite elements in the macroscale problem to complete one iteration/step in the macroscale module. A set of GPU-accelerated solvers may be used for the microscale module to solve multiple RVEs concurrently \cite{mersonNewOpensourceFramework2024}. 

For multiscale analysis involving fibrous material in \mumfim\ implementation, the RVE is a discrete fiber network model such as 3D Delaunay shown in \figref \ref{fig:rve_schematic}b. The material response required at the integration point of a macroscale element is obtained through numerical simulation of the discrete models. Thus, the macroscale procedure does not require any explicit material model of RVE or assumptions regarding the deformation of the fibers. However, as a consequence, the microscale module is computationally expensive for problems of practical interest. The current study aims to develop a surrogate model to replace this module.

To develop the numerical constitutive model, we first systematically sample the deformation gradient landscape and use the GPU-accelerated solvers (microscale module) to produce a large dataset that relates the deformation gradients to corresponding stress and stiffness tensors (i.e., the response of discrete RVEs). We then use a part of the dataset to train an input convex neural network (ICNN) that maps the deformation gradient to strain energy density. The remaining data is used as a test set for model verification purposes. The trained model, along with methods to compute stress and stiffness tensor, forms a complete framework that replaces the microscale module in \figref \ref{fig:rve_schematic}a. The merit is realized as an exceptional gain in computational efficiency with an acceptable margin of error, which is discussed in later sections. 

For the remainder of this section, we describe in greater detail the key aspects of the overall solution procedure, sampling method, and certain choices associated with the neural network framework, ensuring that all interacting parts are compatible and theoretically sound.

\subsection{Multiscale Finite Element Method}
Here, we present the outline of an upscaling multiscale finite analysis scheme that was originally developed by the authors in \cite{mersonNewOpensourceFramework2024}. In our multiscale scheme, the Cauchy stress (\(\vb*{\sigma}\)),  and tangent material stiffness (\(\vout{C}\)) are computed from a set of concurrent subscale problems through computational homogenization.
This can be contrasted against analytical constitutive models that may have been constructed from a priori homogenization, such as the HGO Model \cite{holzapfelNewConstitutiveFramework2000}. The inclusion of this subscale problem is what distinguishes the computational method used here from a run-of-the-mill nonlinear finite element method. This methodology is known as a hierarchical or upscaling multiscale method, where we emphasize that the scheme employed here computes subscale problems concurrently with the macroscale solution procedure.

Two key assumptions in the upscaling multiscale formulation are that the subscale problems have behavior representative of a macroscale continuum and the volume of the subscale problem is of a differential size compared with the macroscale finite element within which it resides. These requirements place strict bounds on the size of the RVE.

\subsubsection{Macroscale Problem}
On the macroscale, we use a 3D updated-Lagrangian displacement-based finite element formulation to solve the balance of virtual power.

\begin{equation}
   \delta P = \delta P^\text{int}-\delta P^\text{ext} = 0 \quad \forall \delta \vb{v} \in \vb*{\mathcal{V}} \label{eq:virtual-power}
\end{equation}
where 
\begin{align}
    \delta P^\text{int} &= \int_V \sigma_{ij} \delta D_{ij} dV, \label{eq:virtual-internal-power} \\
    \delta P^\text{ext} &= \int_{\partial V} t_i \delta v_i \dd{\Gamma}. \label{eq:virtual-external-power}
\end{align}
Here, $\vb*{\mathcal{V}}$ is the set of kinematically admissible velocity fields, $\vb*{\sigma}$ is the Cauchy stress, $\vb{D}$ is the symmetric rate of deformation tensor, $\vb{t}$ is the surface traction, $V$ is the total volume of the body, and $\partial V$ is the surface of the domain.

By pulling out the virtual velocities from the integrals and noting the arbitrariness of the virtual velocities, we obtain the typical nonlinear discretized equations for the residual:
\begin{equation}
    \mathcal{R}_{kI} = \int_V B_{ijkI} \sigma_{ij}  \dd{V}- \int_{\partial V_j} N_{I} t_{k} \dd{\Gamma_j} = 0,
    \label{eq:force-residual}
\end{equation}
and the tangent stiffness:
\begin{equation}
\begin{split}
    K_{kIrJ} = &\int_V B_{ijkI} {\vout{D}}_{ijpq} B_{pq rJ} \dd{V} \\
    + \delta_{ij} &\int_V B_{ijkI} \sigma_{pq} B_{pqrJ} \dd{V},
    \label{eq:tangent-stiffness}
\end{split}
\end{equation}
where majuscule indices correspond to nodal degrees of freedom, and minuscule indices correspond to spatial dimensions. \(\vb{B}_I\) is the shape function derivatives at node \(I\), \(N_I\) is the shape functions at node \(I\), and \(\vout{D}\) is the tangent material stiffness. For the updated-Lagrangian finite element procedure, the material stiffness is given as the push forward of the derivative of the second Piola-Kirchhoff with respect to the Green-Lagrange strain. That is:
\begin{equation}
{\vout{D}}_{mnpq} = \frac{1}{J} F_{mi} F_{nj} \pdv{\Pi_{ij}}{E_{rs}} F_{pr} F_{qs}.
\label{eq:stiffness-push-forward}
\end{equation}

\subsubsection{Downscaling}
An upscaling homogenization scheme requires information passing between the macro and subscale model at each integration point. Downscaling is the process of passing the macroscopic strains \emph{down} to the subscale. Here, we take the subscale displacements as a Taylor series about the macroscopic integration point, that is:
\begin{equation}
    \vb{x}^{\text{s}}\left(\vb{X}^{\text{M}},\vb{X}^{\text{s}} \right) = \vb{X}^{\text{M}} +  \vb{F}^{\text{M}} \cdot (\vb{X}^{\text{s}}-\vb{X}^{\text{M}}) + \Tilde{\vb{x}}^{\text{s}}(\vb{X}^{\text{M}},\vb{X}^{\text{s}}),
    \label{eq:micro_coords}
\end{equation}
where \(\vb{x}^\text{s}\) is the subscale coordinate, \(\vb{F}^\text{M}\) is the macroscopic deformation gradient at the integration point, and \(\Tilde{\vb{x}}^{\text{s}}\) are the higher-order terms that are typically described as fluctuations.
To proceed, we make the following choices, which can be interpreted as orthogonality of the mean and fluctuating fields \cite{luscherEssentialFeaturesFine2012,waltersConsideringComputationalSpeed2021}.
\begin{enumerate}
    \item The RVE centroid is coincident with the macroscopic integration point:
    \begin{equation}
        \int_{V^\text{s}} (\vb{X}^{\text{s}}-\vb{X}^{\text{M}}) \dd{V} = 0.
    \end{equation}
    \item The macroscopic deformation gradient is the volume average of the subscale deformation gradient:
    \begin{equation}
        \vb{F}^\text{M} = \frac{1}{V^\text{s}} \int_{V^\text{s}} \vb{F}^\text{s} \dd{V}.
    \end{equation}
\end{enumerate}

Three types of boundary conditions are common: homogeneous displacement, homogeneous traction, and periodic. 
However, any set of boundary conditions that satisfies:
\begin{equation}
   \int_{\Gamma^{\text{s}}}\Tilde{\vb{x}}^{\text{s}} \otimes \vb{n}\dd{\Gamma} = \vb{0}
\end{equation}
can be used. In this work, we use homogeneous displacement boundary conditions.

Many authors argue for the use of periodic boundary conditions because it reduces the size of the RVE that is needed. However, fibrous materials are not periodic, so this choice is not appropriate. Instead, one may use the so-called "Generalized Boundary Condition'' of \cite{glugeGeneralizedBoundaryConditions2013}, which has been extended to fibrous materials \cite{mersonSizeEffectsRandom2020}. This boundary condition can be used to help mitigate the significant size effect that is present in these materials \cite{mersonSizeEffectsRandom2020}. 

\subsubsection{Upscaling}
Upscaling is the process of transferring data back to the macroscopic problem from the subscale problem. It requires application of the Hill-Mandel criterion, which states that the volume average of an increment in the microscopic power density is equal to the macroscopic power density. Or:
\begin{equation}
    \frac{1}{V^{\text{s}}}\int_{V^\text{s}} \vb*{\sigma}^\text{s} : \delta \vb{D}^\text{s} \dd{V} = \vb*{\sigma}^\text{M} : \delta \vb{D}^\text{M}.
    \label{eq:hill-mandel}
\end{equation}
Based on equation \eqref{eq:hill-mandel}, we identify the following upscaling rule for the stress:
\begin{equation}
    \vb*{\sigma}^\text{M} = 
    \frac{1}{V^{\text{s}}}\int_{V^\text{s}} \vb*{\sigma}^\text{s}\dd{V}.
\end{equation}

An additional rule is needed to upscale the material stiffness as needed for the macroscale tangent stiffness matrix calculation (equation \eqref{eq:tangent-stiffness}). Determination of this upscaling rule is challenging, as it requires a derivative of the mean subscale stresses with respect to the macroscale strains.

If Jacobians are available, one can use the reduced stiffness method described in \cite{kouznetsovaApproachMicromacroModeling2001}, or the non-affine approach described in \cite{mahutgaNonaffineFiberNetwork2023}. However, fiber simulations typically have significant parts of the load path that have singular Jacobians due to nonlinearities. We have found that use of these methods is not robust when large numbers (tens of thousands to millions) of subscale problems must be computed simultaneously. This is because failed convergence of any single RVE prevents the overall macroscale solution from proceeding.

Instead, we make use of an alternative, finite difference approach that works well with the Jacobian free dynamic relaxation solution procedures we use to robustly solve the subscale network problems (see \ref{sec:subscale}). Making use of symmetries does require six additional solves of the subscale problem; however, these typically converge quickly as they only represent small perturbations.

Here, we provide the key equations for computing the material stiffness with the finite difference method; however, a more complete discussion can be found in \cite{mersonNewOpensourceFramework2024}.

From equation \eqref{eq:stiffness-push-forward}, we note that we must find the derivative of the macroscopic second Piola-Kirchhoff stress with the macroscopic Green-Lagrange strain and perform the push forward to obtain \(\vb*{\vout{D}}\), which is used in the updated-Lagrangian finite element calculations. In the remainder of this section, we will use square brackets to refer to the Mandel form of symmetric fourth-order tensor quantities.

The finite difference computation proceeds as follows:
\begin{enumerate}
    \item Compute the stresses in six perturbed directions and compute the finite difference tensor:
    \begin{equation}
    P_{ijpq} = \frac{\Pi_{ij}(U_{lm}+h T_{lmpq})- \Pi_{ij}(U_{lm})}{h},
\end{equation}
where the directions are given by:
\begin{equation}
    \begin{bmatrix}\vb{T}\end{bmatrix} = \frac{1}{2}\begin{bmatrix}
        1 & 0 & 0 & 0& \sqrt{2} & \sqrt{2} \\
        0 & 1 & 0 & \sqrt{2}& 0 & \sqrt{2} \\
        0 & 0 & 1 & \sqrt{2} & \sqrt{2} & 0 \\
        0 & 0& 0& 2& 0& 0 \\
        0 & 0& 0& 0& 2& 0 \\
        0 & 0& 0& 0& 0& 2
    \end{bmatrix},
\end{equation}
in Mandel form.
\item Solve:
\begin{equation}
    \left(\begin{bmatrix}\vb{M}\end{bmatrix}\begin{bmatrix}\vb{T}\end{bmatrix} \right)^T \begin{bmatrix}\pdv{\vb{\Pi}}{\vb{E}}\end{bmatrix}^T = \begin{bmatrix}\vb{P} \end{bmatrix}^T
\end{equation}
where
\begin{align}
   &\begin{bmatrix}
       \vb{M}
   \end{bmatrix} = \nonumber\\ &\frac{1}{2}\left[\begin{smallmatrix}
   2U_{11} & 0 & 0& 0 & \sqrt{2} U_{13} & \sqrt{2} U_{12} \\
   0 & 2U_{22} & 0 & \sqrt{2} U_{23} & 0 & \sqrt{2} U_{12} \\
   0 & 0 & 2U_{33} & \sqrt{2} U_{23} & \sqrt{2} U_{13} & 0 \\
   0 & \sqrt{2} U_{23} & \sqrt{2} U_{23} & U_{22} + U_{33} & U_{12} & U_{13} \\
   \sqrt{2}U_{13} &0 & \sqrt{2} U_{13} & U_{12} & U_{11}+U_{33} & U_{23} \\
   \sqrt{2} U_{12} & \sqrt{2} U_{12} & 0 & U_{13} & U_{23} & U_{11} + U_{22}
   \end{smallmatrix}\right],
\end{align}
and \(U\) is the right stretch tensor.
\end{enumerate}
As described in \cite{mersonNewOpensourceFramework2024}, the probing directions \(\begin{bmatrix}\vb{T}\end{bmatrix}\) are not unique; however, they should correspond to positive definite (tensor) directions.

\subsubsection{Subscale} \label{sec:subscale}
\paragraph{Network Generation}
In this work, we use Delaunay fiber networks \rev{to model the RVE}. The network construction procedure proceeds as follows: First, a generation box is seeded with points using a uniform random distribution. A Delaunay triangularization is then constructed from the initial seed points. To avoid boundary effects, the generation box is trimmed so that the new edge length is half of the one used for the generation box. New nodes are inserted at the intersection of the trimming box and any crossing fibers. The fiber segments that remain outside the trimming box are deleted. This process will yield fiber networks with an average connectivity of \(z\approx14\). This is well above the isostaticity limit for 3D truss networks. The constitutive model for each fiber is given by:
\begin{equation}
    \vb{\Pi} = E \vb*{\varepsilon},
\end{equation}
where $\vb{\Pi}$ is the second Piola-Kirchhoff stress, $E$ is the elastic modulus, and $\vb{\varepsilon}$ is the Green strain. This can alternatively be represented as the force along a fiber \(P = E A_0 \lambda_1\varepsilon_{11}\), where $A_0$ is the initial fiber area.

\paragraph{Solution Procedure}
Due to the difficulties in converging discrete fiber network models, many authors make use of explicit techniques to overcome large material nonlinearities \cite{islamStochasticContinuumModel2018,deogekarStrengthRandomFiber2018}. In these papers, large strains are achieved using dynamic explicit simulations and maintaining the proportion of kinetic energy to strain energy at below five percent. This convergence difficulty was also observed with the Delaunay networks used herein. However, adherence to the Hill-Mandel criterion means that there can be no remaining kinetic energy in the subscale system. To overcome this, we use the dynamic relaxation procedure that was developed in \cite{mersonNewOpensourceFramework2024}. The dynamic relaxation solver makes use of the insight that when a damped dynamic system comes to rest in a state where the internal and external forces (excluding damping) come into balance, this is a state of static equilibrium. Here, a time-dependent central difference solver is used with linear velocity damping (\(\vb{f}^{\text{damp}} = c \vb{v}\)). The damping coefficient and density are parameters that can be chosen to accelerate the convergence.

\begin{figure}
    \centering
    \includegraphics[width = \linewidth]{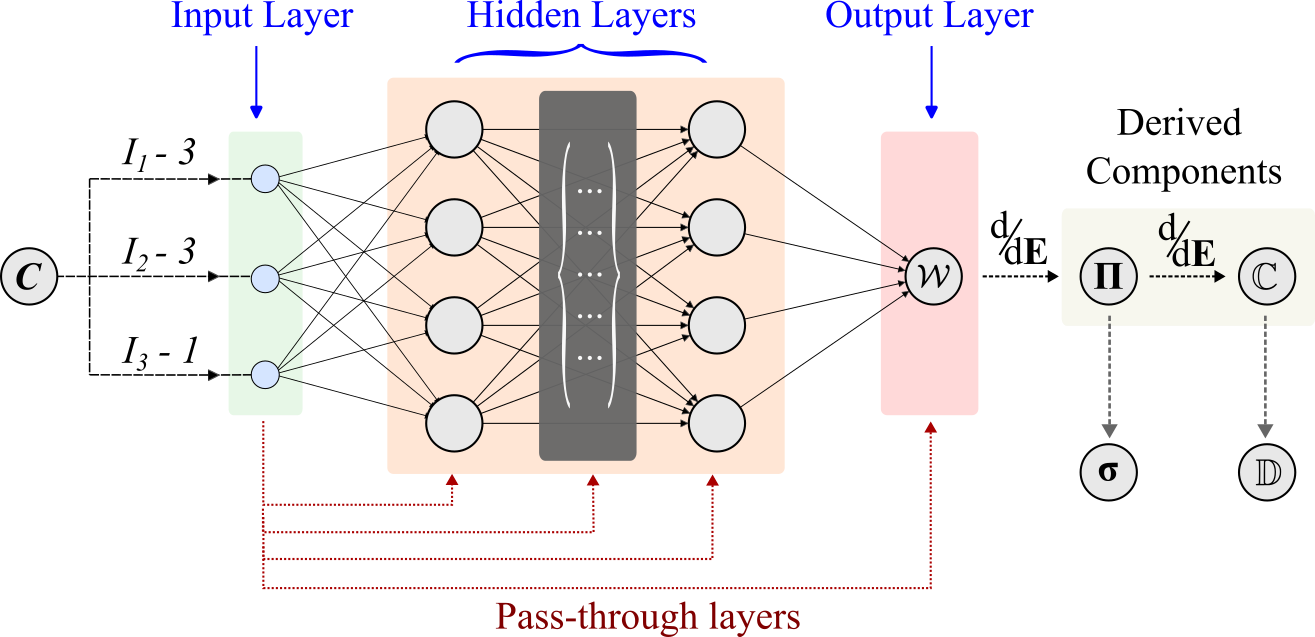}
    \caption{Schematic of the input convex neural network we implement. The input vector consists of invariants of the right Cauchy-Green deformation tensor (\(\srcg\)), and the output is the predicted strain energy density (\(\seng\)). \rev{The components of the PK2 stress (\(\vb{\Pi}\)) and tangent stiffness tensor (\(\vcal{C}=\pdv{\vb{\Pi}}{\vb{E}}\)) are computed by back-propagation through the ICNN neural network. Cauchy stress \(\sigma\) and the updated-Lagrangian material stiffness are needed in the finite element implementation and are computed through the appropriate push forward operations.}}
    \label{fig:icnn_schematic}
\end{figure}

\subsection{Machine Learning Model}
\subsubsection{\rev{Model Selection}}
The majority of the computational cost in the multiscale method is incurred in the distributed parallel analysis of RVEs, owing to the lack of an analytical constitutive model \cite{mersonNewOpensourceFramework2024}. The primary contribution of this work is toward providing a neural network-based framework that will approximately model the fibrous network material. Therefore, the trained model can act as a drop-in replacement of subscale simulations, which reduces the \FET\ method to the traditional FE algorithm that executes at a fraction of the computational cost of the \mumfim \cite{mersonNewOpensourceFramework2024}. To this end, we assume the material behavior is hyperelastic up to the point of failure, which is supported by numerical and experimental results \cite{holzapfelNewConstitutiveFramework2000,parvezStiffeningMechanismsStochastic2023, picuToughnessNetworkMaterials2023}.

The hyperelastic formulation entails finding a non-negative, non-decreasing, and convex scalar energy density function (\(\seng\)) in terms of deformation-related quantities \cite{songHyperelasticContinuumModels2022, tikenogullariDatadrivenHyperelasticityPart2023}. The stress and stiffness tensor components are then found through appropriate first and second-directional derivatives of \(\seng\). The convexity condition implies the machine learning algorithm must be chosen or designed appropriately \cite{amosInputConvexNeural2017, lindenNeuralNetworksMeet2023}. Here, we employ the Input Convex Neural Network (ICNN) model, which guarantees that the overall mathematical form of the neural network will be convex for \rev{appropriate} choice of input and output quantities. The ICNN network was originally pioneered in \cite{amosInputConvexNeural2017}. Various authors have used the ICNN as a basis for hyperelastic material modeling \cite{lindenNeuralNetworksMeet2023, thakolkaranNNEUCLIDDeeplearningHyperelasticity2022, tikenogullariDatadrivenHyperelasticityPart2023, tacDatadrivenModelingMechanical2022, linkaNewFamilyConstitutive2023}. 
\rev{However, this choice is not unique. \citeauthor{tacBenchmarkingPhysicsinformedFrameworks2024} provides an excellent benchmark between Constitutive Artificial Neural Networks (CANN), ICNNs, and Neural Ordinary Differential Equations, finding that they all exhibit significant limitations while being used to fit a skin dataset \cite{tacBenchmarkingPhysicsinformedFrameworks2024}. They additionally show that their trained networks exhibit large variations in the stiffness, further emphasizing the need to incorporate stiffness into the training dataset. Here, we chose to use the ICNN as we found our downstream finite element models were more likely to converge with the ICNN based network rather than the CANN network described in \cite{linkaNewFamilyConstitutive2023}.} A schematic of the neural network model we implemented is shown in \figref \ref{fig:icnn_schematic}.

\subsubsection{\rev{Neural Network Architecture}}
\rev{The output of a fully connected feed-forward neural network is a convex function of its input if the $i$-th layer of the neural network is a component-wise convex function of $i-1$-th layer \cite{kleinPolyconvexAnisotropicHyperelasticity2022}. ICNN is a particular form of fully connected feed-forward neural network with restrictions on activation functions and weights to satisfy this criterion.} The convexity of the model is ensured by strictly adhering to (i) convex activation functions and (ii) non-negative weights. \rev{From an implementation point of view, the first requirement is trivial and entails using convex activation functions such as \texttt{softplus} \cite{softplus_ref,kleinPolyconvexAnisotropicHyperelasticity2022}}. The second requirement, i.e., non-negative weights, limits the scope of the neural network when modeling complex phenomena \cite{amosInputConvexNeural2017}. To counter, a number of pass-through connections without restrictions on weights or biases are introduced that connect the inputs to the \rev{hidden and output layer neurons}. These pass-through connections amount to an affine mapping of input \rev{and thus do not affect the convexity of the model}. 


\rev{In this study, we aim to train the ICNN to obtain a strain energy density function that is polyconvex in terms of the applied deformation, e.g., $\seng = \seng(\vb{F}, \text{cof }\vb{F}, \text{det }\vb{F})$ or $\seng = \seng(I_1, I_2, I_3)$. Here, $I_1, I_2$, and $I_3$ correspond to the invariants of the right Cauchy-Green deformation tensor with $(I_1, I_2, I_3)$ being convex in $(\vb{F}, \text{cof }\vb{F}, \text{det }\vb{F})$ respectively \cite{kleinPolyconvexAnisotropicHyperelasticity2022}. Following past works, e.g., \cite{linkaNewFamilyConstitutive2023, thakolkaranNNEUCLIDDeeplearningHyperelasticity2022}}, the input layer of ICNN consists of modified invariants of the right Cauchy-Green deformation tensor ($I_1 - 3$, $I_2 - 3$, and $I_3 - 1$). This choice helps to normalize the input values. We employ the \texttt{softplus} activation function defined as $g(x) = \dfrac{1}{\beta} \log(1 + \beta \exp(x))$ with $\beta = 1$, a choice inspired by \cite{thakolkaranNNEUCLIDDeeplearningHyperelasticity2022, lindenNeuralNetworksMeet2023, kleinPolyconvexAnisotropicHyperelasticity2022}. The function is non-negative, non-decreasing, and convex in $x$. A second function, $f$, of similar characteristics, is applied over the weights to fulfill the requirements for positive weights. Here, we choose $f = g$ without loss of generality. Therefore, the resulting general form of the strain energy density function is non-negative, non-decreasing, and convex function \rev{of the invariants} by construction. 

This model, however, does not guarantee vanishing strain energy at zero strain (i.e., \(\seng|_{\vb{F}=\vb{I}}\)) due to the nature of activation functions. \rev{Specifically, the \rev{\texttt{softplus}} activation function we use produces an output equals $g(0) = \log(2) > 0$ at zero input, which leads to \(\seng|_{\vb{F}=\vb{I}} > 0\)}. The optimization algorithm can overcome this shortcoming by adjusting the biases in the pass-through layers. Figure \ref{fig:pred_energy} indicates that the requirement is approximately satisfied. An alternative approach can be introducing standalone bias terms to act as correction factors, for example, in \cite{thakolkaranNNEUCLIDDeeplearningHyperelasticity2022}.

\rev{Finally, we note that the choice of input is dependent on the underlying material model and available dataset. For the isotropic and elastic fibrous material we model, our input array choice is appropriate. For anisotropic materials, additional quantities describing the anisotropy are suggested, e.g., \cite{LINKA2023116007}. Similarly, some invariants such as $I_3$ can be ignored if certain assumptions, such as incompressibility, are applicable.} \revtwo{A brief discussion of the impact of removing the $I_2$ and $I_3$ invariants is presented in Section \ref{sec:constitutive_model}.}

\begin{figure}
    \centering
    \includegraphics[width = \linewidth]{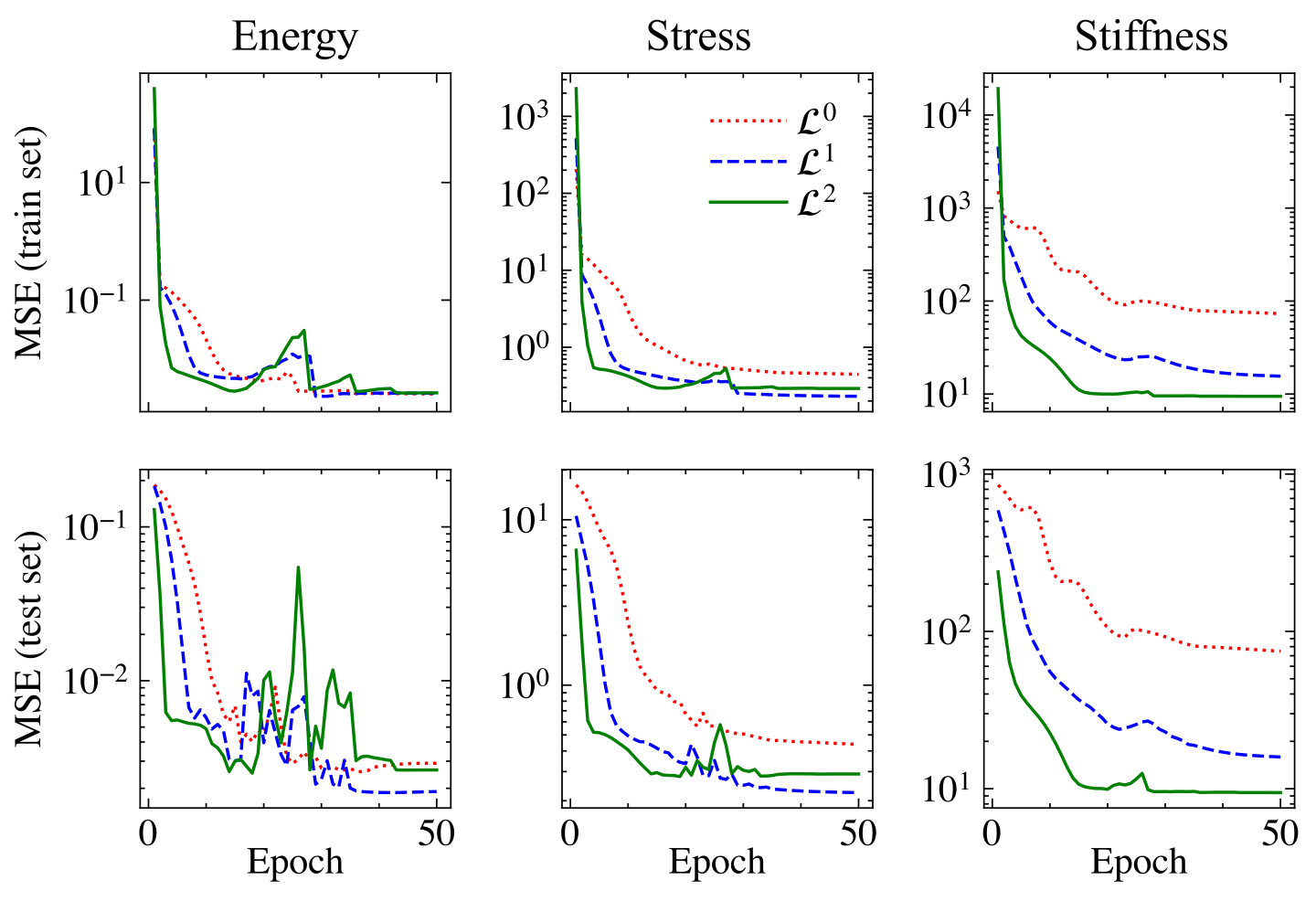}
    \caption{Mean squared error (loss) of ICNN during training. The average errors associated with strain energy density, stress, and stiffness tensors are shown in the first, second, and third columns, respectively. The top row shows the loss for the training dataset, and the bottom row represents the corresponding results for the test dataset. The legend \(\slossk{k} := \sum_{i=0}^{i=k} H_i\) indicates the variant of the loss function used to train ICNN, with \(k\) being the highest order of derivative included in the composite loss function. \revtwo{The input array consists of all isotropic invariants.}}
    \label{fig:icnn_mse}
\end{figure}

\subsubsection{ICNN Implementation}
The ICNN we train contains five fully connected hidden layers, each having four neurons in the primary branch. The pass-through layers connect the input vector to these hidden layer neurons. Mathematically, the output of the neurons in the $i$-th hidden layer is $\vb{z}^{(i)} = g(f(\vb{A}^{(i)}) \vb{z}^{(i-1)} + \vb{A}^{(i)}_p \vb{x} + \vb{b}^{(i)}_p)$. Here, $\vb{x}$ is the input vector to the model, and $A^{(i)}$ is the weight matrix for the $i$-th layer of the primary branch. The final two terms ($A^{(i)}_p$ and $b^{(i)}_p$) represent the weight matrix and bias vector due to the pass-through connection to the $i$-th layer, respectively. The model is \rev{trained and evaluated over} 91,590 samples, with an 80/20 split for the train and test dataset. We use Adams solver for parameter optimization without any weight decay \cite{kingmaAdamMethodStochastic2017}. The model parameters are initialized using Kaiming uniform distribution \cite{heDelvingDeepRectifiers2015}. We use the mean squared error $\text{MSE}(y) = (y_{\text{pred}} - y_{\text{true}})^2$ as the template for the loss function.  

In the original and most subsequent implementations of ICNN, the loss function is based only on the output variable. For our purpose, this translates to hyperparameter optimization based on strain energy density. However, due to the large number of trainable parameters, the prediction accuracy for energy does not generally translate to its derivative quantities, such as stress and stiffness, which are necessary for FE simulations. To increase the prediction accuracy of the derivatives, we implement the so-called Sobolev training protocol such that the model is trained on the target output and its derivatives concurrently \cite{czarneckiSobolevTrainingNeural}. Here, we include both the first and second derivatives in the ICNN training procedure as follows-
\rev{
\begin{enumerate}
    \item For the \(i\)-th batch iteration within an epoch, we compute the following average losses. The stress and stiffness components are found through automatic differentiation of ICNN updated at ($i-1$)\textsuperscript{th}.
    \begin{enumerate}
        \item Loss in energy, $H_0 = \text{MSE}(\seng)$
        \item Loss in stress, \(H_1 = \langle \text{MSE}(\sstr)\rangle\) (averaged over the components)
        \item Loss in stiffness, \(H_2 = \langle \text{MSE}(\sstiff) \rangle\) (averaged over the components)
    \end{enumerate}
    \item The total loss per batch is computed as \(\slossk{k} = \sum_{j=0}^{j=k}\rev{\gamma_j} H_j\), with $k=2$ being the highest order of derivative considered. Here, \(\gamma_j\) is a weighting factor for each term. We find \(\gamma_j=1 \; \forall j\in [0,k]\) to be adequate despite the relative difference in units and magnitudes of $ H_j$.
    \item The model parameters are updated based on the total loss, and the cycle is repeated.
    \item The learning rate is adjusted automatically at the end of an epoch based on the mean loss. The learning rate decreases by a factor of ten if the \rev{model exhibits non-decreasing} loss for five consecutive epochs.
\end{enumerate}
}
We use a moderately large batch size (\rev{128}), so the loss (especially in the gradients) approximated from a batch is representative of the population. Finally, we note that the construction of neural network determines the convexity and smoothness of the model. Sobolev training is an add-on that greatly improves the prediction accuracy of higher-order quantities. 

\revtwo{Sobolev training requires accurate first and second derivative calculations. To compute them, we follow the following procedure starting with the assumption that the underlying material behavior is hyperelastic (has a strain energy density function) which leads to the definition of the second Piola-Kirchhoff as \(\Pi = \pdv{\seng{}}{\vb{E}} =  2\pdv{\seng{}}{\vb{C}}\), where \(\vb{C}\) is the right Cauchy-Green deformation tensor. And, the material stiffness is given as \(\vcal{C}=\pdv{\vb{\Pi}}{\vb{E}} = 2\pdv{\vb{\Pi}}{\vb{C}}\). The standard push forward operations are used to transform the stress and stiffness to the reported measures that correspond to those used by \mumfim{}'s updated-Lagrangian finite element solution procedure. The push forward for the stiffness tensor is given in equation \eqref{eq:stiffness-push-forward}.}

\revtwo{
These derivatives of the strain energy density, \(\seng\), are computed through backpropagation and make use of the PyTorch auto differentiation capabilities. Although we use the invariants of the right Cauchy-Green deformation to restrict ourselves to an isotropic material model, they are computed using PyTorch vectors, which allows us to perform backpropagation to the right Cauchy-Green deformation tensor. Alternatively, we could have computed the derivatives with respect to the invariants and used the standard relations to obtain the stiffness and stress.}

\revtwo{
The performance of the backpropagation, especially for second derivatives, was significantly impacted by the implementation details. Details of these methods and ICNN implementations are provided in our GitHub repository \footnote{https://github.com/LACES-LAB/ML-biotissues/blob/main/imports/fibernn/derivatives.py} and SI.
}


\subsubsection{Training Data Generation}
One of the major challenges with developing machine learning models is that it is difficult to know if a given input was reasonably covered by the training dataset. We have developed a novel sampling methodology that allows for explicit control over the range of principal stretch values in which we may be interested. This explicit control provides a mechanism to verify whether a given deformation state is likely to be covered by the training data.

Since our constitutive response must be frame-indifferent, we can sample the set of positive definite right stretch tensors. The standard relations can be used to map the resulting stress and stiffness back to a rotated frame as needed for the analysis procedures.

We employ the following algorithm to construct our set of positive definite right stretch tensors:
\begin{enumerate}
    \item Sample principal stretches \(\lambda_1, \lambda_2, \lambda_3 \in [\lambda_{\text{min}}=0.85,\lambda_{\text{max}}=1.15]\), using Latin hypercube sampling \label{sample_start}
    \item Construct base right stretch tensor: \begin{equation}
        \vb{U}_i^* = \begin{pmatrix}
            \lambda_1 & 0 & 0 \\
            0 & \lambda_2 & 0 \\
            0 & 0 & \lambda_3
        \end{pmatrix}
    \end{equation}
    \item Sample a random rotation matrix: \(\vb{R}_j\) \label{sample_R}
    \item Construct a realization of the stretch tensor using the polar decomposition formula:
    \begin{equation}
       \vb{U}_j = \vb{R}_i \vb{U}_i^* \vb{R}_j^T
    \end{equation} \label{sample_Ui}
    \item Repeat steps \ref{sample_R}--\ref{sample_Ui} until sufficient samples are collected for this set of stretches \label{sample_repeat1}.
    \item Repeat steps \ref{sample_start}--\ref{sample_repeat1} until a sufficient number of base stretches are sampled.
\end{enumerate}

If a constant number of samples is used within each ellipsoidal shell that is represented by a given base stretch \(\vb{U}^*_i\), then each data point will cover a varying volume in phase space. To avoid this problem, the number of rotation samples constructed for each set of base stretches is scaled by the surface area of the ellipsoid, with \(\lambda_1, \lambda_2, \lambda_3\) forming the ellipsoid axes. This will ensure that, in a statistical sense, each sampled point will cover a constant volume in phase space, which is critical to avoid oversampling and associated overfitting. These sampled deformations are used as inputs to our GPU accelerated network solver in \mumfim{} to generate the neural network training and test dataset. \rev{Some key characteristics of the generated dataset are provided in SI.}

\subsubsection{Model Availability}
The neural network model was implemented using the PyTorch \cite{paszkePyTorchImperativeStyle2019} framework and is available for public use along with the dataset on which the network was trained.\footnote{The dataset is available at https://zenodo.org/doi/10.5281/zenodo.11205879} \footnote{The source code is available at https://github.com/LACES-LAB/ML-biotissues} The detailed numerical values of model parameters omitted here for brevity are available in the code repository. 

\begin{figure*}
    \centering
    \includegraphics[width = \linewidth]{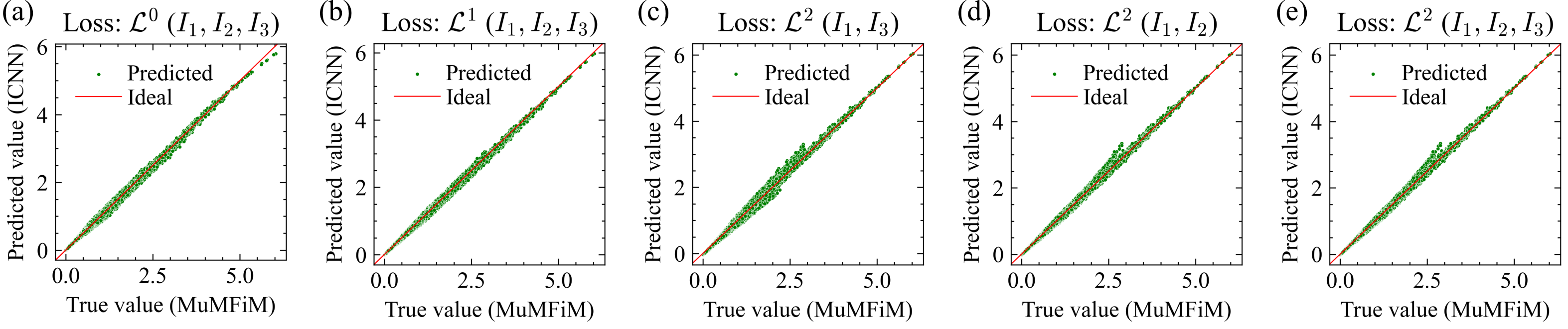}
    \caption{True (\mumfim) vs ICNN-predicted strain energy density for various deformation gradients ($F$) in test dataset using ICNN trained with $\slossk{2}$ loss function. The dashed red line indicates the ideal fit, and scatters indicate predictions.}
    \label{fig:pred_energy}
\end{figure*}

\section{Results and Discussions}
In this section, we first present the model's predictive capability and the improvements brought forth by the Sobolev training protocol. We then present a case study involving mechanical deformation of the \rev{FCL}, simulated using the trained model. Finally, we compare the output to the results of \mumfim\ based large-scale numerical simulation to comment on the efficacy of the current approach.

\subsection{Sobolev Training of ICNN}
We first discuss the overall predictive accuracy of the input convex neural network (ICNN) in terms of the average loss of desired quantities. This energy, stress, and stiffness result is shown in the first, second, and third columns of \figref \ref{fig:icnn_mse}. We compare three variants of the model emanating from the choice of the composite loss function (i.e., $k = 0, 1, 2$ in $\slossk{k} = \sum_{j=0}^{j=k} \rev{\gamma_j} H_j$), indicating the highest derivatives included in the loss function.

The first column of \figref \ref{fig:icnn_mse} shows the loss in \textit{strain energy density} for different choices of the overall loss function. Expectedly, the choice of loss function has a negligible effect on the predictive accuracy of strain energy density. The derivative quantities (such as stress), however, benefit from including $H_1$ in the loss function (curves corresponding to \(\slossk{1}\)). This result is shown in the second column of \figref \ref{fig:icnn_mse}. The additional information passed to the optimizer also improved the predictive accuracy of stiffness, as shown in the final column of the figure. Naturally, including the loss in stiffness in the overall loss function (curves corresponding to \(\slossk{2}\)) improved the predictive accuracy of the stiffness for both the training and test datasets. 

Comparing the loss function variants, it is evident that including higher-order derivatives of the output function improved the accuracy of the derivative quantities at any given epoch. The substantial improvement in stiffness prediction accuracy is especially beneficial for the downstream finite element algorithm, which would be a typical use case of the ICNN-based model developed here. 

\rev{In fact, we found that the model variant with \(\slossk{2}\) led to a more robust finite element solution that was able to converge with a wider range solver parameters. 
And, the accuracy of the stresses was lower in the \(\slossk{0}\)  and \(\slossk{1}\) cases. This impact is particularly noticable outside of range of the training data which can be seen in the context of a single RVE in \figref \ref{fig:RVE_stress_strain}.} Consequently, subsequent discussions primarily focus on the \(\slossk{2}\) variant only. 

\subsection{Numerical Constitutive Model}\label{sec:constitutive_model}
At this point, we comment on the accuracy of the trained ICNN for individual components of the desired outputs. We use the normalized MSE, computed as \(\text{NMSE}(y) = \text{MSE}(y)/\text{mean}(y_{\text{true}}^2)\), as the measure of deviation from the true value (numerical solution).

\revtwo{Here, we present the true vs predicted values of the strain energy density for the test samples in \figref \ref{fig:pred_energy} for various choices of the loss function and input invariants}. In this figure, the ICNN-predicted strain energy density corresponding to the numerical solution is shown via scatters. The red line indicates the ideal match between true and predicted value (i.e., slope = 1.) The computed NMSE for strain energy density is approximately 0.15\%, indicating exceptional predictive accuracy. \revtwo{Further, the accuracy of the strain energy is similar for all choices of loss function, and input invariants.}

\begin{figure*}
    \centering
    \includegraphics[width = \linewidth]{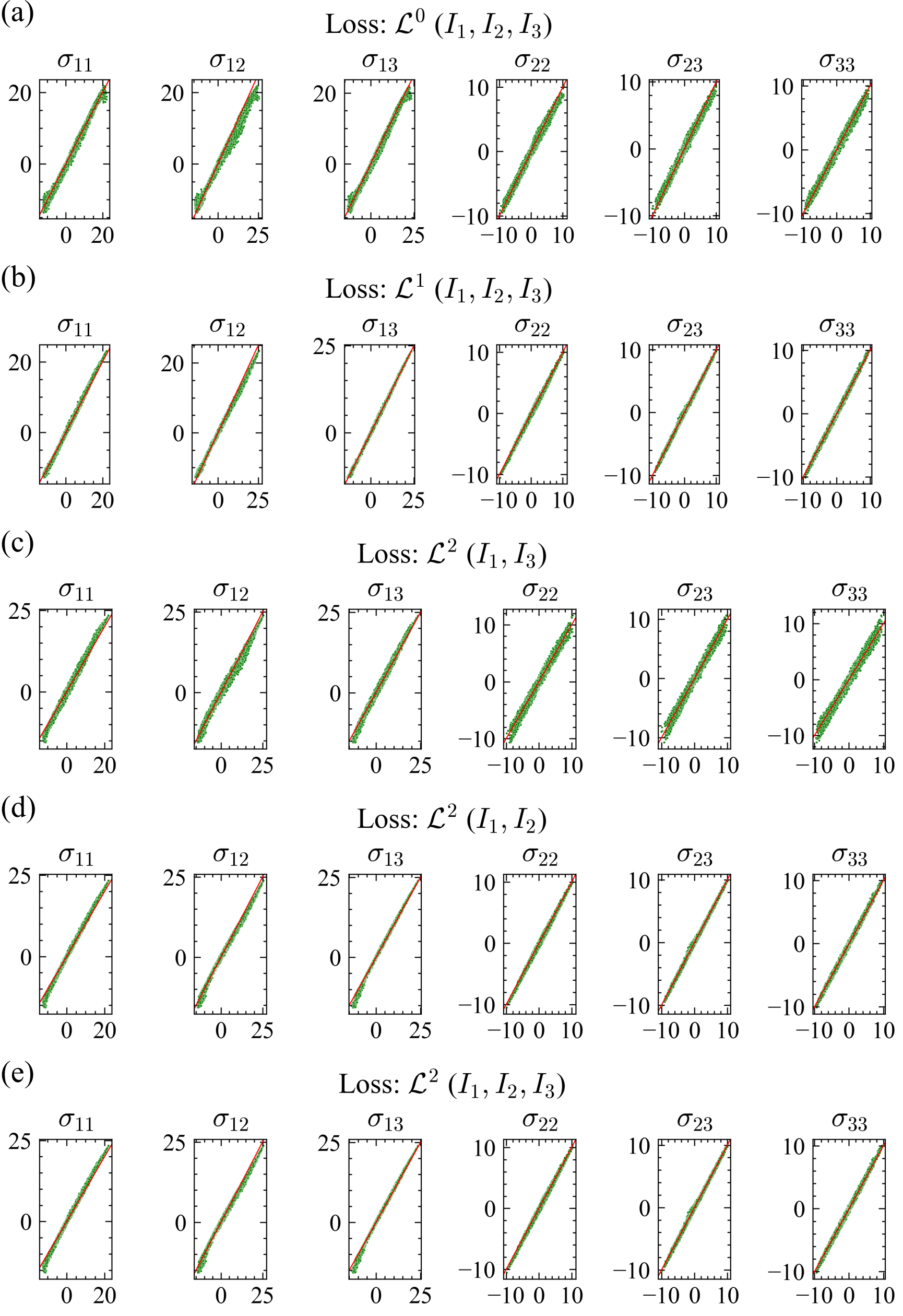}
    \caption{True (\mumfim) vs ICNN-predicted values for unique components of Cauchy stress tensor for various deformation gradients ($F$) in the test dataset. \revtwo{The variant of the loss function used and invariants of $\vb{C}$ included in the input array are indicated in sub-figure titles.}}
    \label{fig:pred_stress}
\end{figure*}

\revtwo{The corresponding results for the unique components of Cauchy stress tensors are presented in \figref \ref{fig:pred_stress} which reaffirms the conclusion drawn from \figref \ref{fig:icnn_mse}, viz, the predictive accuracy for stresses is similar for $\slossk{1}$ and $\slossk{2}$ when the input array consists of all isotropic invariants. The average NMSE for the stress tensor components is 0.815\%, with a maximum of 1.25\% from \(\sigma_{22}\) component with $\slossk{2} (I_1, I_2, I_3)$.}

\revtwo{
However, unlike energy, omitting $I_2$ as an input degrades the predictive accuracy for stress for the material under consideration. This is seen as a wide band in \figref \ref{fig:pred_stress}c. Omitting $I_3$ as an input does not impact the predictive accuracy of the stress (\figref \ref{fig:pred_stress}d). This indicates that the material can be considered nearly incompressible. Detailed error measures for the stress components are given in SI (Fig 3).}

\begin{figure*}
    \centering
    \includegraphics[width = \linewidth]{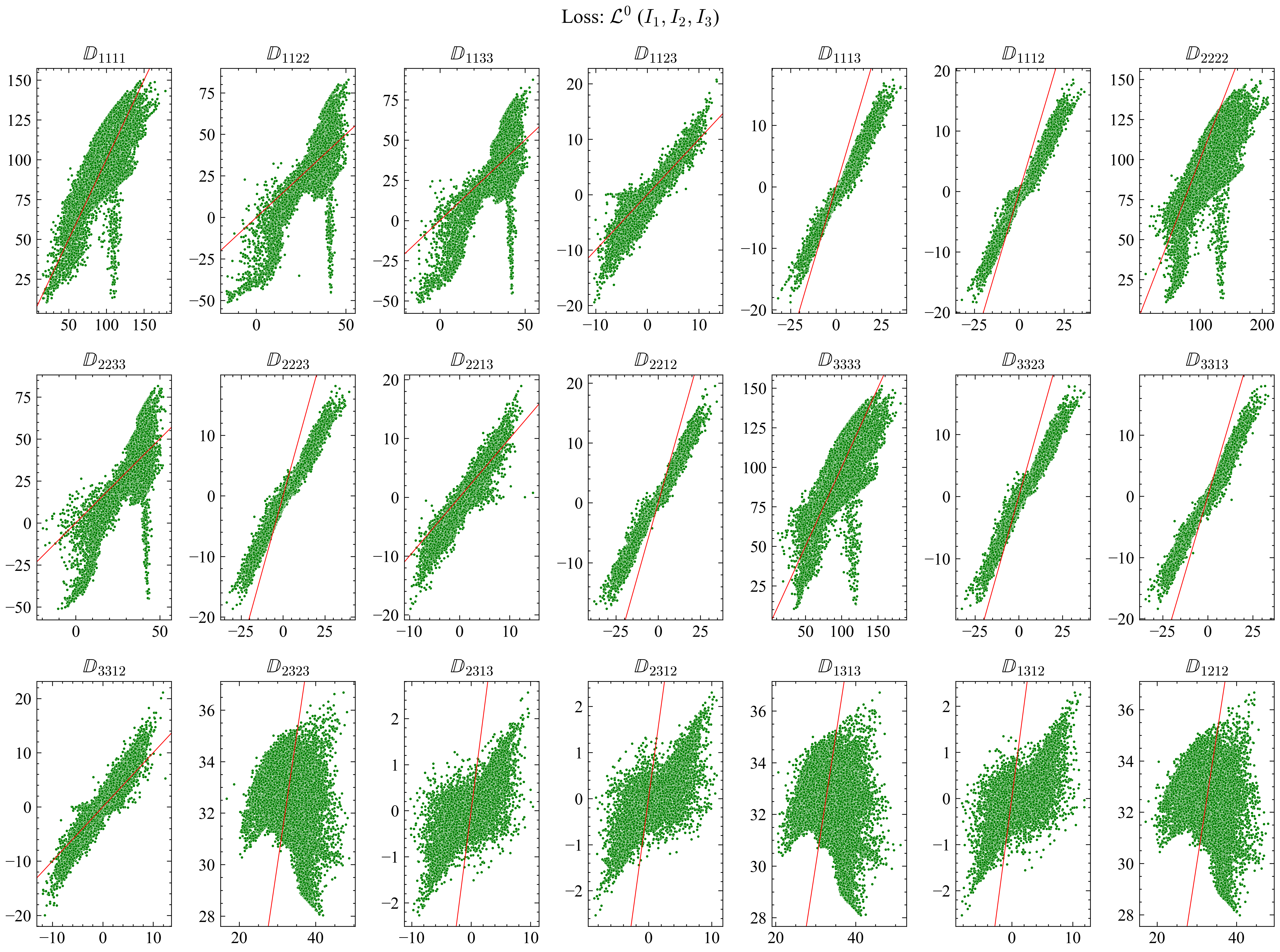}
    \caption{True (\mumfim) vs ICNN-predicted values of the unique components of stiffness tensor for various deformation gradients ($F$) in test dataset using ICNN trained with $\slossk{0}$ loss function. The input array includes all isotropic invariants of $\vb{C}$.}
    \label{fig:pred_stiff_l0}
\end{figure*}

\begin{figure*}
    \centering
    \includegraphics[width = \linewidth]{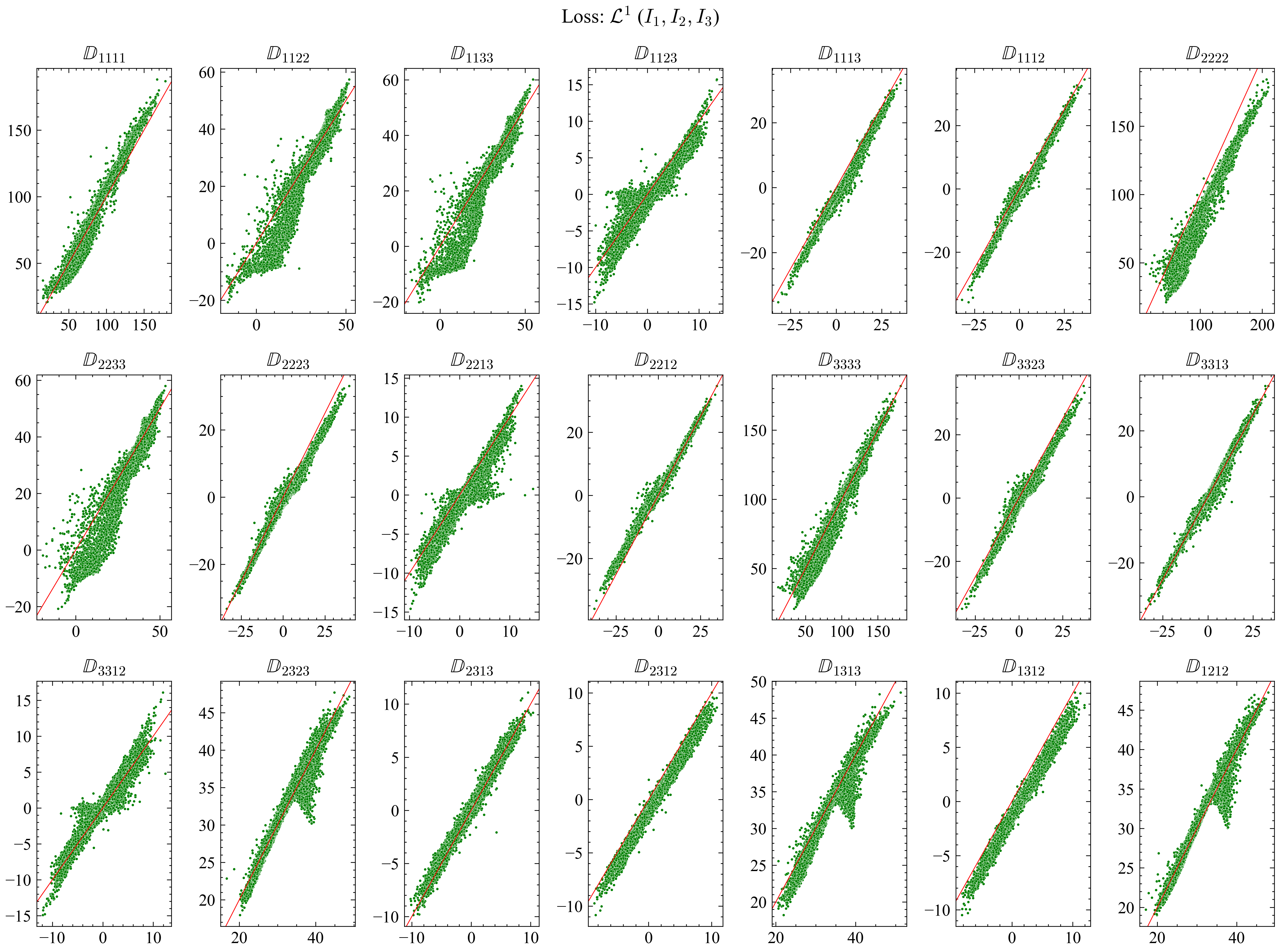}
    \caption{True (\mumfim) vs ICNN-predicted values of the unique components of stiffness tensor for various deformation gradients ($F$) in test dataset using ICNN trained with $\slossk{1}$ loss function. The input array includes all isotropic invariants of $\vb{C}$.}
    \label{fig:pred_stiff_l1}
\end{figure*}

\begin{figure*}
    \centering
    \includegraphics[width = \linewidth]{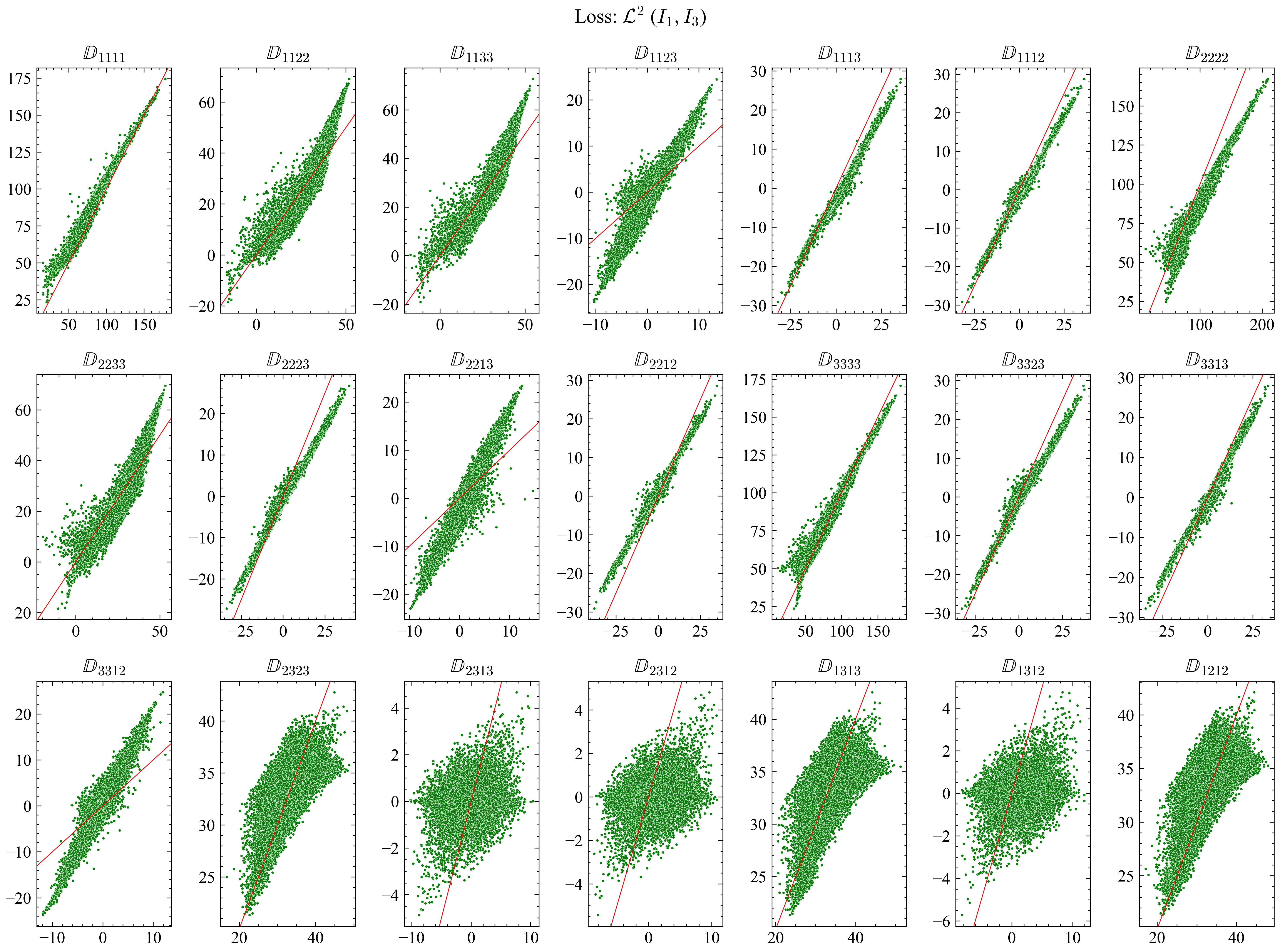}
    \caption{True (\mumfim) vs ICNN-predicted values of the unique components of stiffness tensor for various deformation gradients ($F$) in test dataset using ICNN trained with $\slossk{2}$ loss function with $I_1$ and $I_3$ as input.}
    \label{fig:pred_stiff_l2_i13}
\end{figure*}

\begin{figure*}
    \centering
    \includegraphics[width = \linewidth]{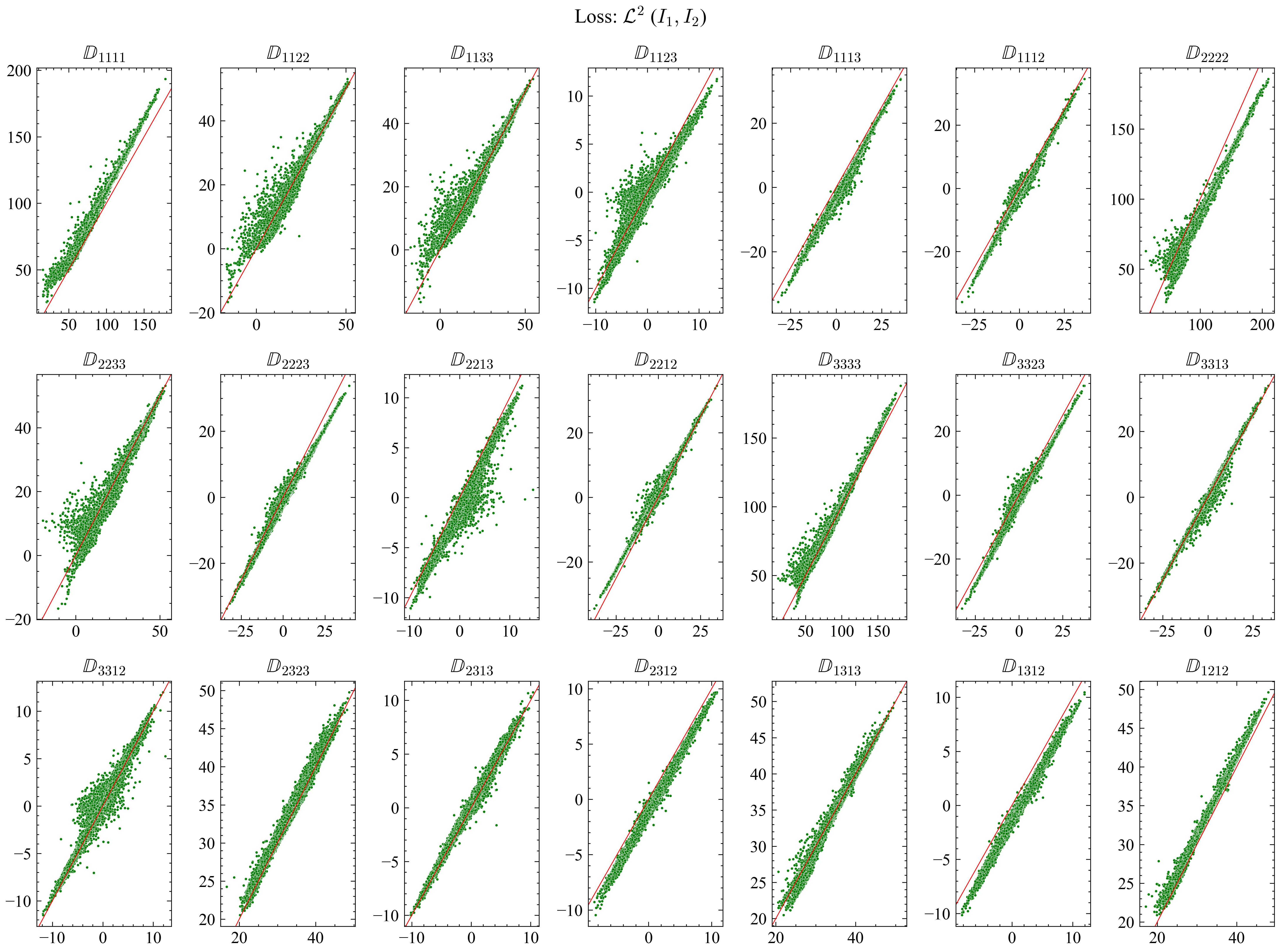}
    \caption{True (\mumfim) vs ICNN-predicted values of the unique components of stiffness tensor for various deformation gradients ($F$) in test dataset using ICNN trained with $\slossk{2}$ loss function with $I_1$ and $I_2$ as input.}
    \label{fig:pred_stiff_l2_i12}
\end{figure*}

\begin{figure*}
    \centering
    \includegraphics[width = \linewidth]{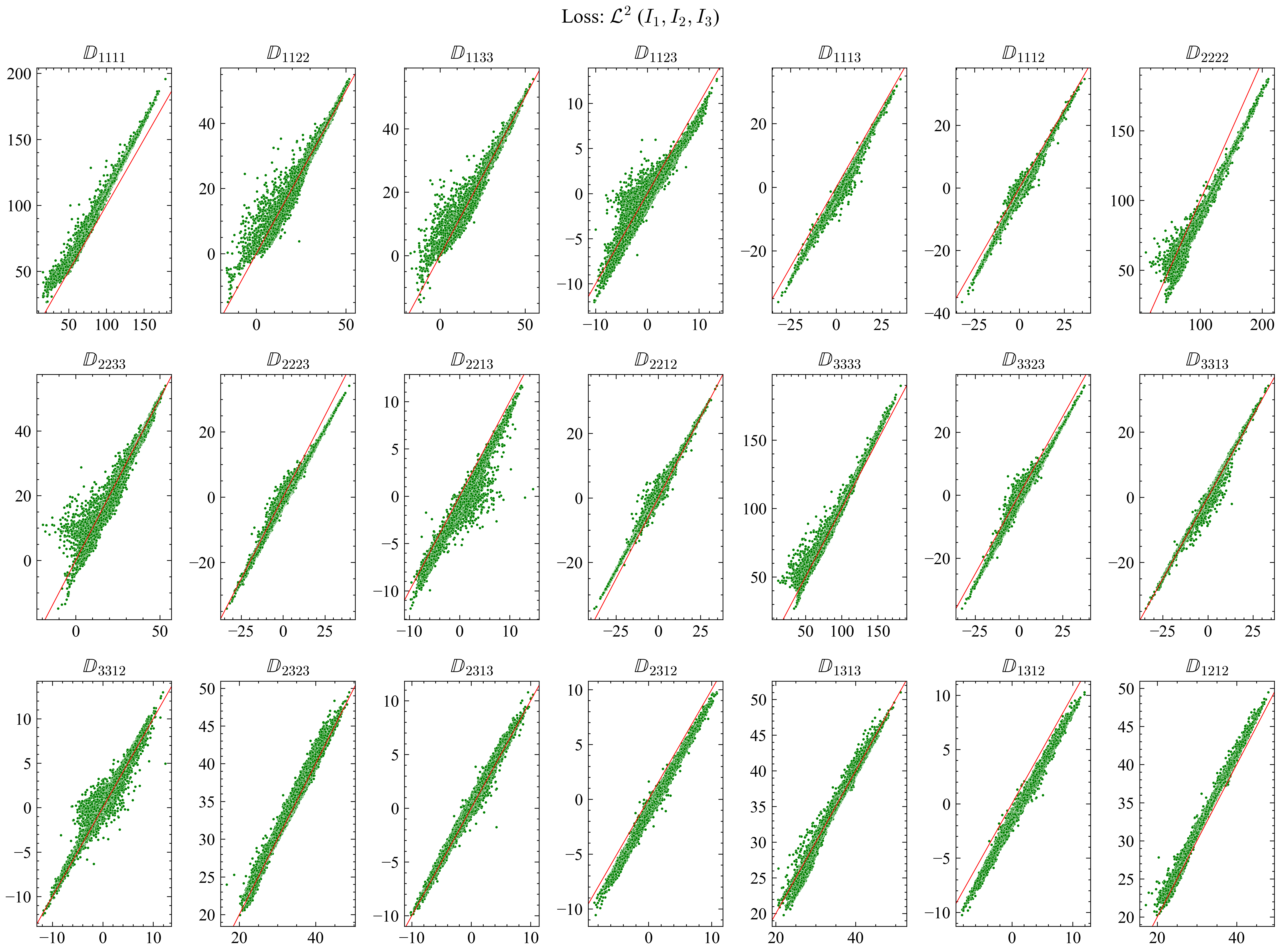}
    \caption{True (\mumfim) vs ICNN-predicted values of the unique components of stiffness tensor for various deformation gradients ($F$) in test dataset using ICNN trained with $\slossk{2}$ loss function. The input array includes all isotropic invariants of $\vb{C}$.}
    \label{fig:pred_stiff_l2}
\end{figure*}

\revtwo{We observe a similar trend for the components of stiffness tensor visualized in \figref \ref{fig:pred_stiff_l0}-\ref{fig:pred_stiff_l2}. For the given ICNN architecture, the prediction accuracy is best of $\slossk{2}$ variant with at least $I_1$ and $I_2$ included in the input array. Here, 
the mean normalized error is 5.4\% for the stiffness tensor components. The maximum deviation is for \(\vout{D}_{1312}\) at 23.35\%, which is also reflected as a visible drift from the ideal fit in \figref \ref{fig:pred_stiff_l2} and \ref{fig:pred_stiff_l2_i12}}.

This loss of accuracy for stiffness is not necessarily surprising. Two potential reasons are \rev{(i) poor signal-to-noise ratio at the small strain limit in the dataset due to higher order derivatives evaluated using numerical methods} and (ii) the near-constant gradient of the activation functions at large inputs. The modeling is further complicated by the fact that the underlying data relates to a stochastic fibrous material, which is not guaranteed to have a smooth energy landscape. Despite these challenges, the machine-learned constitutive model is able to faithfully reproduce results on a single RVE (\figref \ref{fig:RVE_stress_strain}) and on an exemplar \rev{FCL} geometry (\figref \ref{fig:sim_compare}).

\revtwo{Other variants of the model, e.g., $\slossk{0}, \slossk{1}$ and $\slossk{2} (I_1 \& I_3)$ generally performed worse for material under consideration as shown in \figref \ref{fig:pred_stiff_l0}, \ref{fig:pred_stiff_l1}, and \ref{fig:pred_stiff_l2_i13} respectively. The detailed error measures for the variants are available in SI (Fig 4)}.

\subsection{Single RVE Results}

\begin{figure*}
    \centering
    \begin{minipage}[b]{0.3\linewidth}
    \centering
    \includegraphics[width=\linewidth]{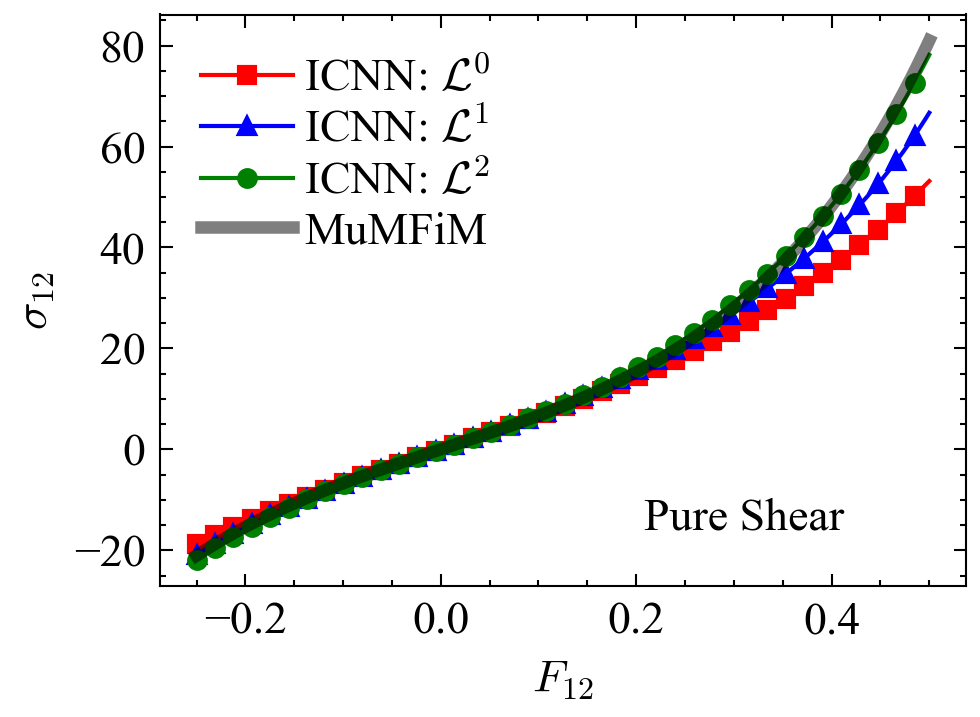}
    \end{minipage}
    \hfill
    \begin{minipage}[b]{0.3\linewidth}
    \centering
    \includegraphics[width=\linewidth]{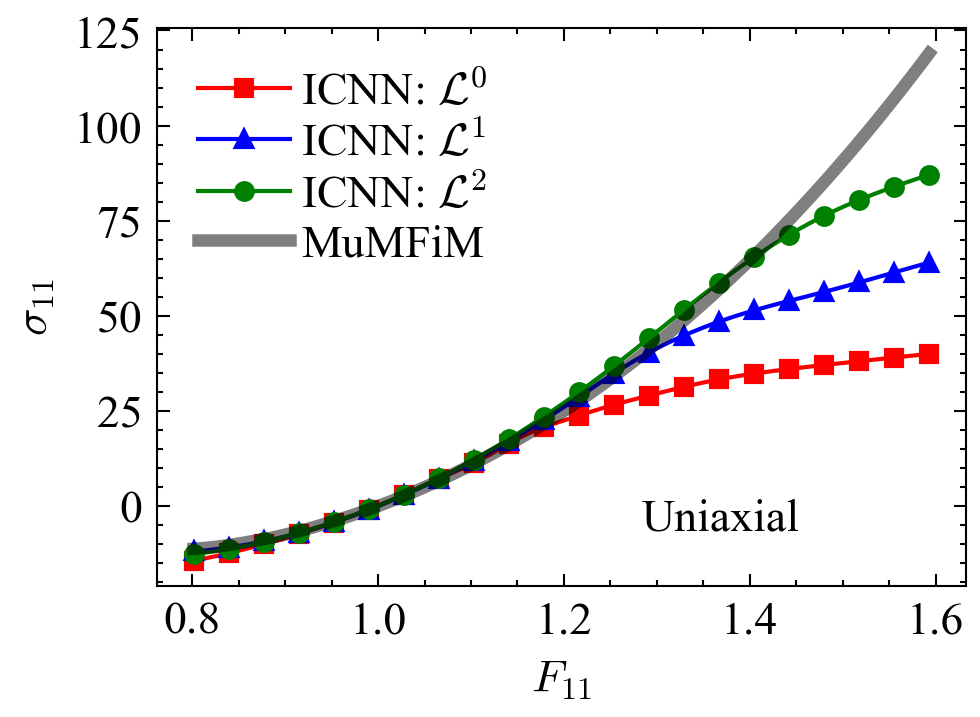}
    \end{minipage}
    \hfill
    \begin{minipage}[b]{0.3\linewidth}
    \centering
    \includegraphics[width=\linewidth]{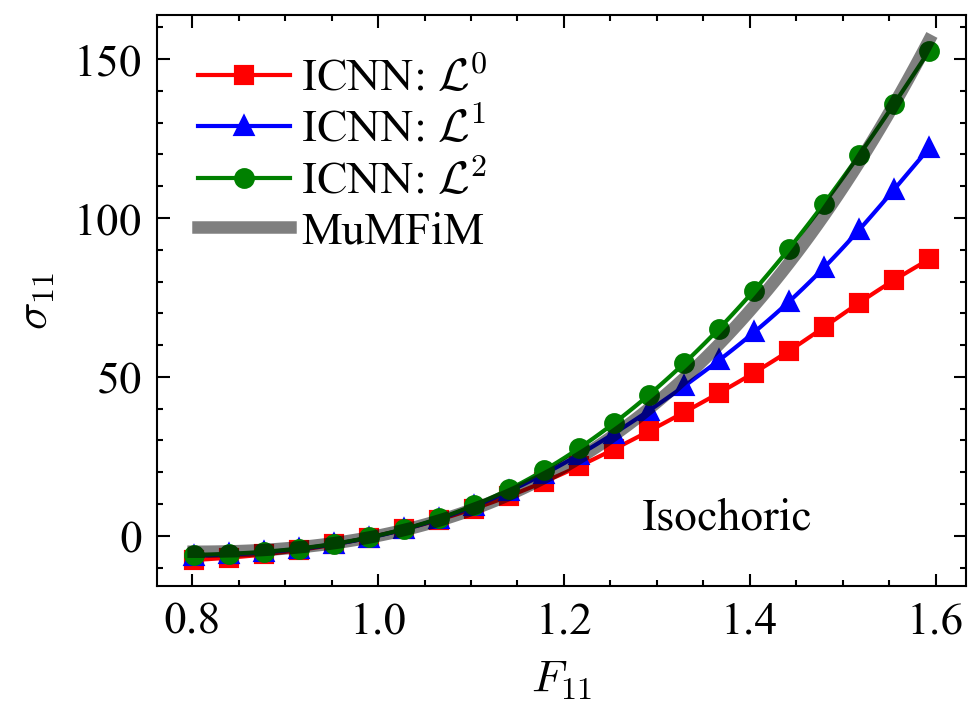}
    \end{minipage}
    \caption{Comparison of Cauchy stress ($\sigma$) vs deformation gradient ($F$) between a single \mumfim{} RVE and ICNN-predicted values for each considered composite loss function undergoing pure shear (left), uniaxial deformation (center), and isochoric deformation (right). For each case, the ICNN neural network with the $\slossk{2}$ composite loss function maintains accuracy outside the training regime ($0.85 \leq \lambda \leq 1.15$). \revtwo{The ICNN was trained with all isotropic invariants of $\vb{C}$}}
    \label{fig:RVE_stress_strain}
\end{figure*}

To understand the applicability of the ICNN-based model across a range of multiaxial loading states, we performed a comparison against a single \mumfim{} RVE shown in \figref \ref{fig:RVE_stress_strain}. We applied homogeneous displacement boundary conditions described with the following three deformation gradients:
\begin{itemize}
    \item Pure Shear
    \begin{equation}
        \vb{F} = \begin{pmatrix}
            1 & \alpha/2 & 0 \\
            \alpha/2 & 1 & 0 \\
            0 & 0 & 1
        \end{pmatrix}
    \end{equation}
    \item Uniaxial
    \begin{equation}
        \vb{F} = \begin{pmatrix}
            1+\alpha & 0 & 0 \\
            0 & 1 & 0 \\
            0 & 0 & 1
        \end{pmatrix}
    \end{equation}
    \item Isochoric
    \begin{equation}
        \vb{F} = \begin{pmatrix}
            1+\alpha & 0 & 0 \\
            0 & 1/\sqrt{1+\alpha} & 0 \\
            0 & 0 & 1/\sqrt{1+\alpha}
        \end{pmatrix}
    \end{equation}
\end{itemize}

We observe that for each case, increasing the number of derivative terms in the composite loss function increases the regime where the ICNN-based model accurately predicts the constitutive response. This is most pronounced in uniaxial extension. We additionally observe that with the $\slossk{2}$ composite loss function, the ICNN-based model maintains accuracy outside the training regime ($0.85 \leq \lambda \leq 1.15$). These results indicate that it is beneficial to include higher than second derivatives into the training process. To further emphasize the utility of the ICNN-based models undergoing complex multiaxial load states, we apply them in a 3D finite element simulation of the \rev{FCL}.

\subsection{Case Study: Uniaxial Tension of Facet Capsular Ligament}
To demonstrate the efficacy of the ICNN-based model, we present a comparison of the results of a boundary value problem with a 3D model of a \rev{FCL}. \rev{The shape of the FCL cross-section was obtained from \cite{banCollagenOrganizationFacet2017} and is representative of a human Cervical FCL. The cross-section is uniformly extruded into the plane with a thickness of 0.5mm using Siemens NX CAD. The bounding box around the ligament is 11.84mm by 8.56mm in the x and z directions. This model was meshed and model attributes were assigned in Simmetrix SimModeler \cite{SimmetrixSimulationModeling}. The mesh has 63,821 linear tetrahedral elements. Meshes were converted to the PUMI native format and model attributes were exported as a model-traits YAML file which are read by \mumfim{} \cite{ibanezPUMIParallelUnstructured2016,mersonModeltraitsModelAttribute2021}.}

\rev{For the case presented herein, Delaunay networks are constructed by the procedure in \ref{sec:subscale} that serve as RVE for \mumfim. Fiber orientations are uniformly distributed in the initial configuration (no preferential alignment). They have average density defined as the total fiber length per unit volume of \(186.12\) and a mean fiber length of \(0.216\).
}

\begin{figure}
    \centering
    \includegraphics[width = \linewidth]{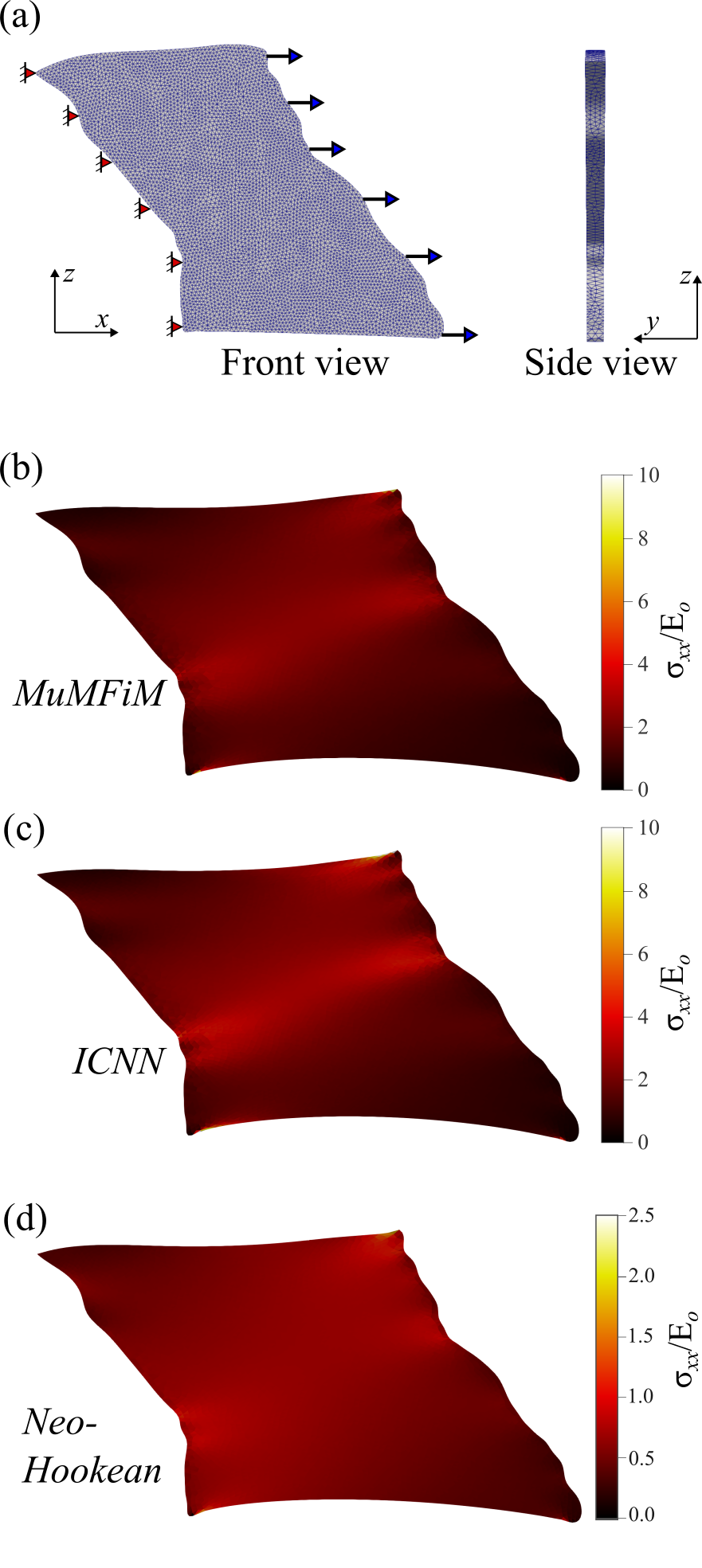}
    \caption{(a) Geometry and boundary conditions for numerical simulation of FCL under uniaxial tension along the x-axis. \rev{This model has 63,821 linear tetrahedral elements.} (b) Deformed configuration at 70\% strain based on large scale \mumfim/\FET\ simulation. (c) Deformed configuration at same strain using standard finite element method with trained ICNN (\(\slossk{2}\) variant) as the constitutive descriptor, and (d) with calibrated neo-Hookean material model. The color indicates the true stress in the elements along the loading direction \rev{normalized by the small strain stiffness of the material, $E_o$}. \rev{A different color bar is used for the Neo-Hookean model to emphasize that both the magnitude and structure of the field are different for the Neo-Hookean material. The reader is referred to \figref{9} for additional insight into the magnitude of the difference.}}
    \label{fig:sim_compare}
\end{figure}

We consider ``bone'' boundary conditions on the left and right surfaces of the FCL model. In this setup, the left surface is held encastre, and the right surface is fixed in the $y$ and $z$ directions and displaced in the $x$ direction. The meshed geometry and boundary conditions are shown schematically in \figref \ref{fig:sim_compare}(a). The remaining boundaries are in a traction-free condition. We solve this problem in two ways: First, we use the \mumfim\ package to obtain a fully numerical solution. Next, we solve the same problem with trained ICNN as the material model. In other words, we replace the 'microscale module' in \figref \ref{fig:rve_schematic}a with an ICNN-based framework. We treat the \mumfim\ solution as the reference result for this comparison. A more detailed study of this problem is available in \cite{mersonNewOpensourceFramework2024}. 

The deformed configurations at approximately 70\% global strain obtained from ICNN and \mumfim{}-based simulations are shown in \figref \ref{fig:sim_compare}b and \ref{fig:sim_compare}c, respectively. The color represents the true stress in the elements along the loading direction. Visually comparing these results, it is immediately clear that the ICNN-based approach closely replicates the predictions of \mumfim, including the shear-band-like phenomena appearing due to the geometry of the ligament. We also show the result obtained \rev{from a} compressible neo-Hookean model \rev{that has been calibrated on the small strain data from \mumfim{}} in \ref{fig:sim_compare}d. This model predicts a more diffused band with a far smaller maximum stress. Although neo-Hookean is the simplest material model for hyperelastic behavior, it captures the essence of the difficulties associated with classical continuum models in the context of fibrous network materials. The small parameter space of classical models and strict assumptions regarding the shape of the strain energy density functions fail to capture the complexities arising from the kinematics of the fibers in fibrous materials. A neural network model such as the one developed here thus presents a trade-off between explainability and generality. 

\begin{figure}
    \centering
    \includegraphics[width = \linewidth]{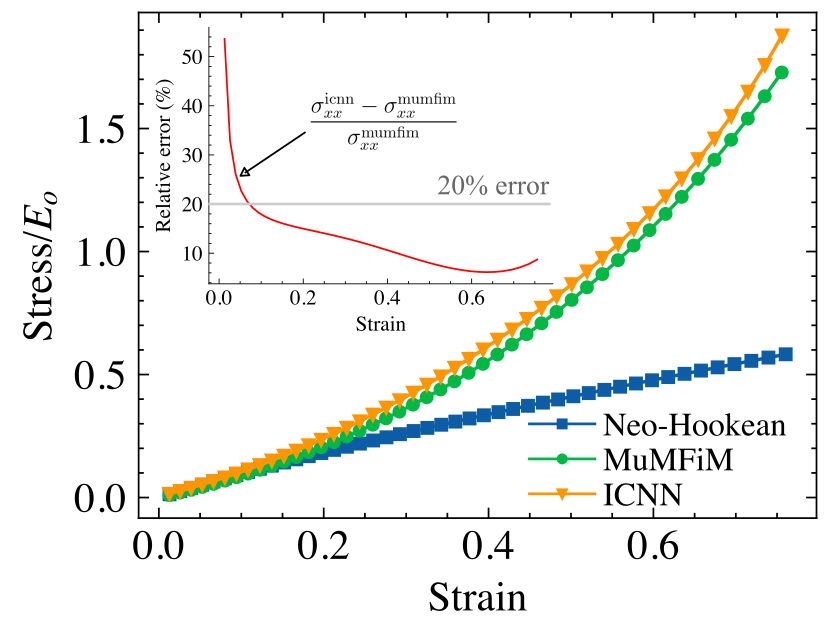}
    \caption{Cauchy stress as a function of Green-Lagrange strain in the loading direction based on \mumfim\, neo-Hookean, and ICNN-based simulations of FCL under uniaxial tension. The relative error of ICNN-based prediction with respect to \mumfim\ output is shown as inset.}
    \label{fig:str_strain}
\end{figure}

The stress-strain response of the \rev{FCL} extracted from the \mumfim\ and ICNN-based simulation are presented in \figref \ref{fig:str_strain}. Both approaches predict exponential stiffening, which is a hallmark behavior of soft biological tissues \cite{islamEffectNetworkArchitecture2018, parvezStiffeningMechanismsStochastic2023, licupStressControlsMechanics2015}. The neo-Hookean model calibrated \rev{with \mumfim{} data at a small strain limit does not show similar exponential stiffening.} The results indicate that the ICNN-based method retains reasonable accuracy up to a very large strain while reducing the computational resource requirement by orders of magnitude—from GPU-accelerated super-computers to consumer-grade laptops. 

The inset of \figref \ref{fig:str_strain} shows the measured relative error of the ICNN result for stress with respect to \mumfim. The maximum relative error is in the small-strain regime and upward of 50\%. The deviation, however, falls rapidly below 20\% at approximately 10\% strain. \rev{The discrepancy in the small strain regime is likely related to a similar deviation in the predicted stiffness at the small strain limit (\figref \ref{fig:pred_stiff_l2}). We suspect the application of numerical noise at this limit is behind this discrepancy}. In fibrous network materials such as biological tissues, however, the large strain regime is the principal concern \cite{picuNetworkMaterialsStructure2022}. Given the computational cost for large-scale simulation of fibrous materials and the technical difficulties required to develop efficient algorithms such as those employed in \cite{mersonNewOpensourceFramework2024}, the current model presents an excellent pathway for obtaining approximate results involving the deformation of fibrous materials. \rev{To put the cost savings into perspective, the multiscale analysis requires one RVE solve per element in each iteration of the macroscale solver. In this case, we used a line-search method that does three evaluations to compute a valid search distance with an average of six iterations per load step. For the 60 load steps in the FCL model, \mumfim{} must perform nearly 69 million evaluations of the discrete fiber network RVEs (63821 elements * 6 line search per step * 3 iterations per line search * 60 steps). Unsurprisingly, these RVE evaluations account for the majority of the \mumfim{} solve time. The dataset that was used to train the model in this paper only required 91,961 evaluations of the discrete RVE.}
Although our work does not obviate the need for large-scale numerical simulations, it does provide a pathway for those in the community to make use of results from those expensive simulations with modest computational resources.

\subsection{Limitations}
The ICNN-based model developed here suffers from several limitations. The major shortcomings are enumerated below.
\begin{enumerate}
    \item \textit{Limited availability of training data:} Our model relies on the higher-order derivatives of the energy function to achieve superior predictive accuracy for quantities such as stiffness tensor components. These datasets are generally not readily available, especially for novel materials. Under these limiting circumstances, we suggest \rev{employing $\slossk{0}$ or $\slossk{1}$ loss function variant or models described in \cite{linkaNewFamilyConstitutive2023, thakolkaranNNEUCLIDDeeplearningHyperelasticity2022}} while being mindful of the loss of accuracy for higher-order quantities.
    \item \textit{Limited accuracy in the small strain regime}: Our model does not enforce $\seng = 0$ for $\vb{F} = \vb{I}$ due to the model architecture, specifically, the same constraints that ensure the model convexity. However, the optimizer can adjust the bias in pass-through layers without any constraints to push $\seng$ arbitrarily close to zero. The trained model indeed shows such convergence for strain energy density and stress (\figref \ref{fig:pred_energy} and \ref{fig:pred_stress}). \rev{We also observe relatively poor predictive accuracy of stiffness tensor components at the small strain limit, which may be a cause of concern for specific use cases}.
    \item \textit{Anticipated loss of accuracy for extreme deformation:} We anticipate the model will gradually lose accuracy beyond the scope of the training dataset. This is particularly important if the problem of interest contains singularities such as sharp cracks. In such use cases, the predictions should be scrutinized carefully, or the model should be re-trained with an extended dataset if necessary. 
\end{enumerate}

We hope the published code and dataset will increase interest in this approach and lead to more general models encompassing other microstructural aspects, such as fiber alignment and flocculation. 

\section{Conclusion}
In this paper, we have developed a novel numerical constitutive model based on an Input Convex Neural Network (ICNN) that is able to accurately predict energies, stresses, and stiffnesses for fibrous materials without limiting assumptions such as affine deformation of microscale fibers or incompressibility. This neural network preserves important constitutive constraints such as polyconvexity and frame indifference through its structure. The strict enforcement of these constraints is critical to obtaining outputs that can be used in well-posed finite element simulations. By construction of higher-order tensor quantities through back-propagation of derivatives, we are additionally able to obtain symmetric stiffness tensors as required by the mathematical description of hyperelastic materials.

We show that constitutive constraints alone are not enough to construct accurate models of stress and stiffness. The contributions of this work stem from: First, the incorporation of the Sobolev training protocol, which involves including higher-order derivatives in the loss function, significantly improves the model's prediction accuracy for stress and stiffness. Second, the trained model, when used as a replacement for the subscale segment of multiscale procedures such as \FET\ or as a material model in standard finite element procedures, provides a quantitatively similar solution to practical problems of interest at a fraction of the computational cost. Despite their accuracy, performing \FET\ multiscale simulations requires significant computational resources that are out of reach for most biomechanical researchers. Our material model provides a bridge between the computational demands of multiscale simulations and the typical accuracy needed to derive important biomechanical insights. We exemplify this through the analysis of a \rev{FCL} and believe that others can benefit from applying this \rev{model and} method to their research.


\bibliography{refs} 


\end{document}


\maketitle
\section{About Dataset}
Here, we provide a few features of the training/test dataset produced using the algorithm outline in ``Section 2.2.1: Training Data Generation'' of the paper. The dataset contains 91590 entries in total. Through random sampling, 80\% of the entries were selected for training purposes and 20\% for testing. Only the deformation gradients, $\vb{F}$, were sampled using the described algorithm. The rest have been generated using standard tools from continuum mechanics formulations, e.g., $\vb{C} = \vb{F}^T \vb{F}$, or MuMFiM simulation of discrete fibrous network RVE with prescribed $\vb{F}$. The distributions of the few relevant quantities are shown below. The dataset is publicly available in the repository specified in the main paper.

\begin{figure}[h!]
    \centering
    \begin{subfigure}[t]{0.3\textwidth}
        \centering
         \includegraphics[width=\textwidth]{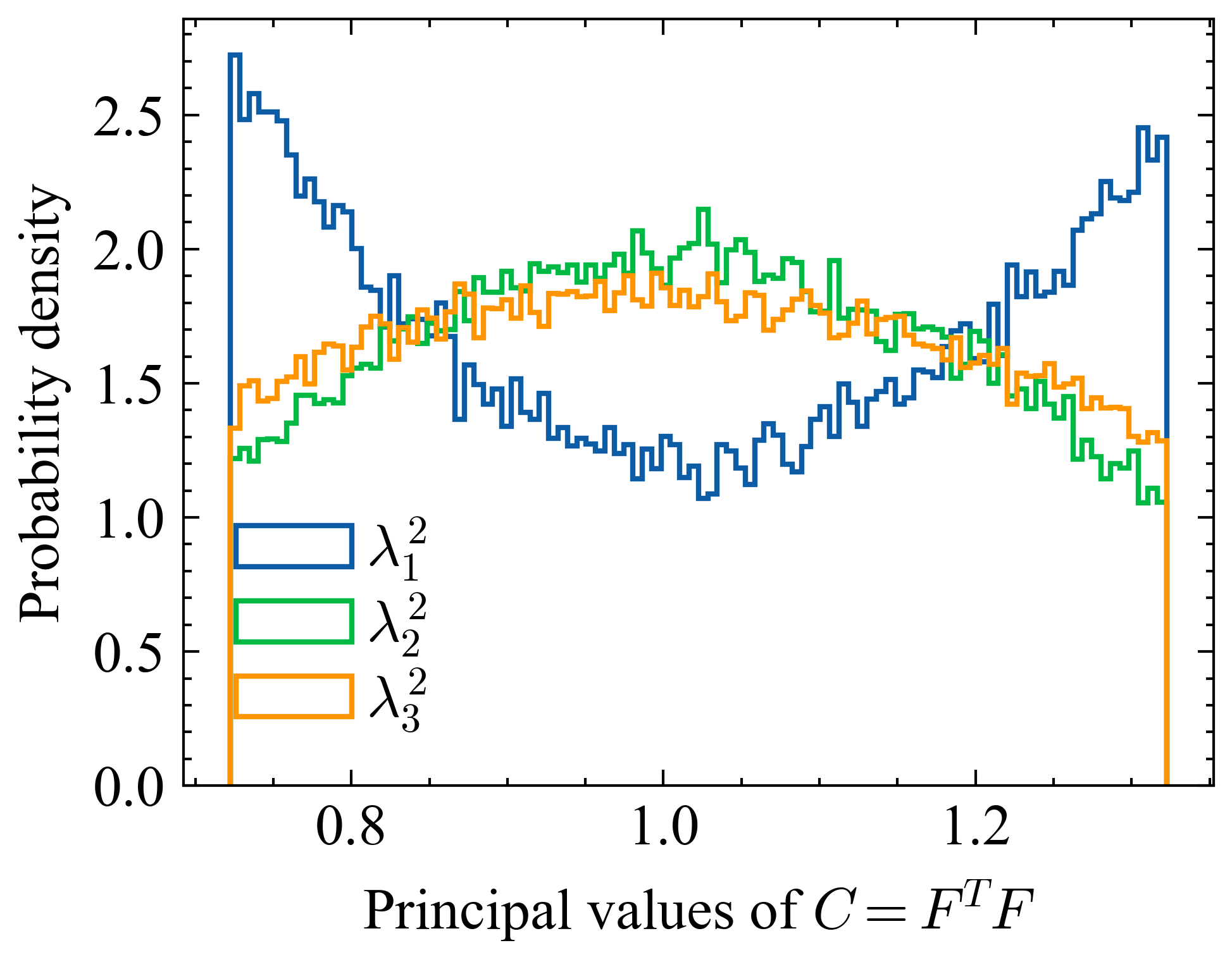}
         \caption{Distribution of principal values of right Cauchy-Green deformation tensor, $\vb{C}$}
    \end{subfigure}
    \hfill
    \begin{subfigure}[t]{0.3\textwidth}
        \centering
         \includegraphics[width=\textwidth]{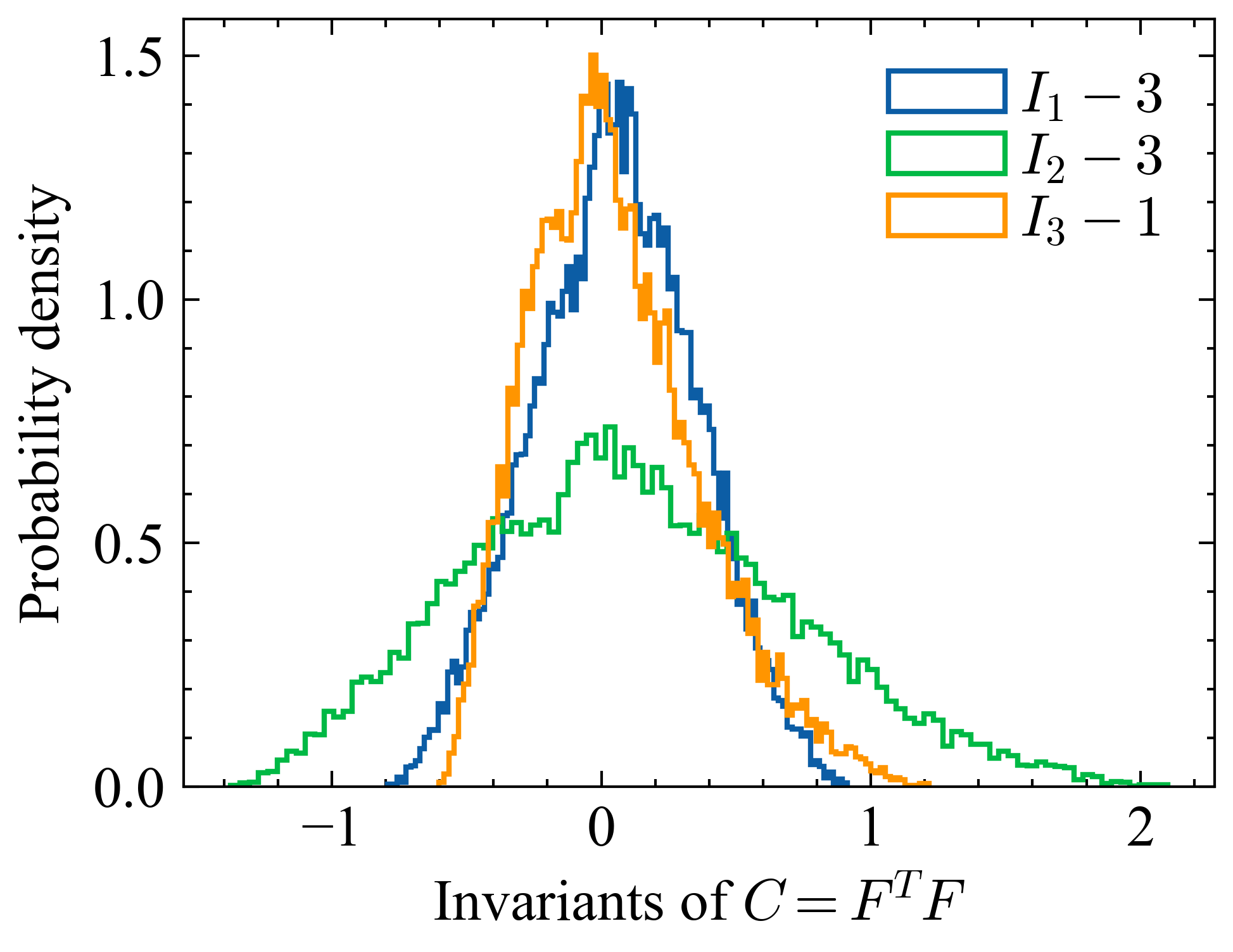}
         \caption{Distribution of invariants of right Cauchy-Green deformation tensor, $\vb{C}$}
    \end{subfigure}
    \hfill
    \begin{subfigure}[t]{0.3\textwidth}
        \centering
         \includegraphics[width=\textwidth]{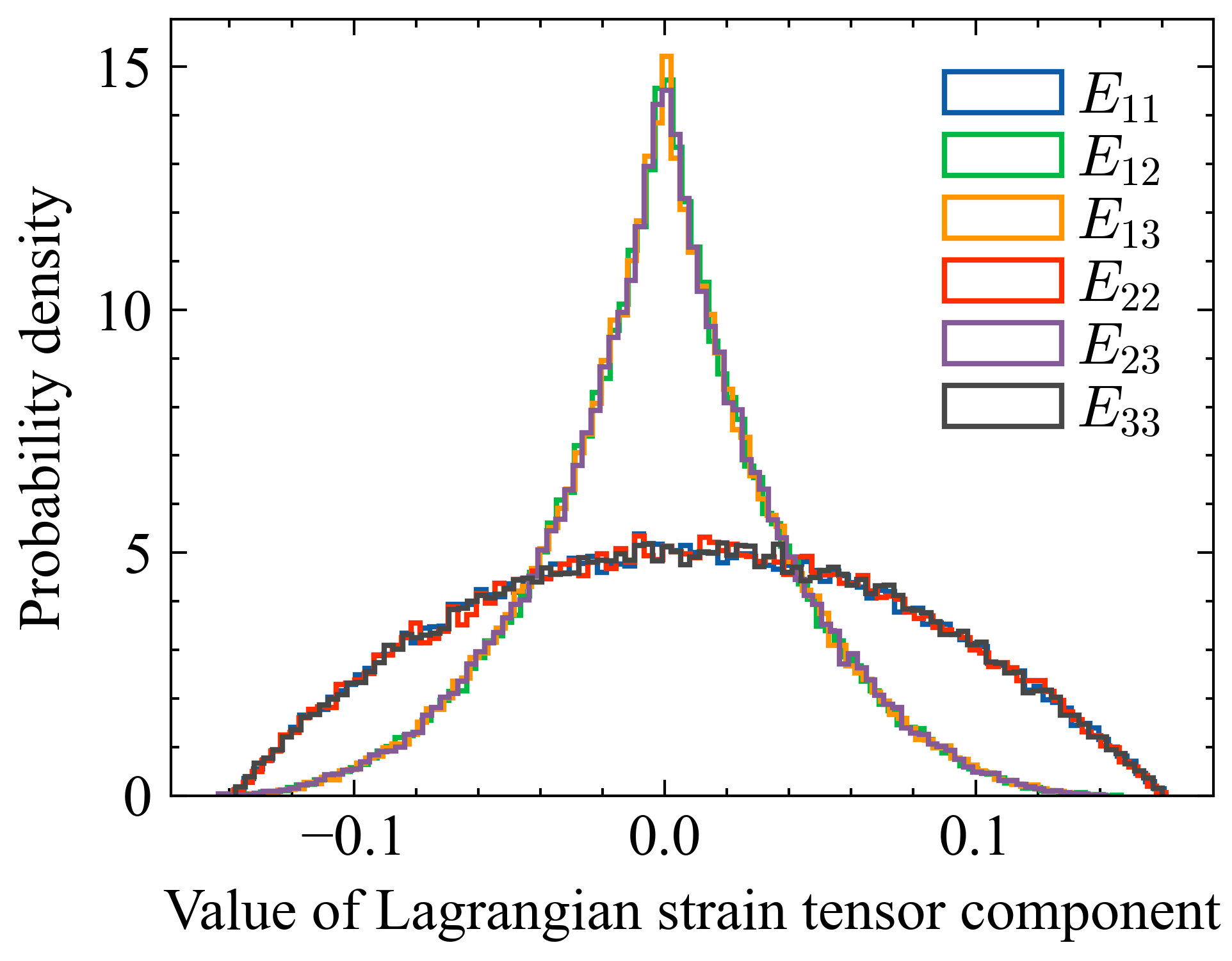}
         \caption{Distribution of entries of Lagrangian strain tensor, $\vb{E}$}
    \end{subfigure}
    
    \caption{Relevant features of sampled deformation gradients, $\vb{F}$ for ICNN training.}
    \label{fig:sampled_F}

\end{figure}

\begin{figure}[h!]
    \centering
    \begin{subfigure}[t]{0.3\textwidth}
        \centering
         \includegraphics[width=\textwidth]{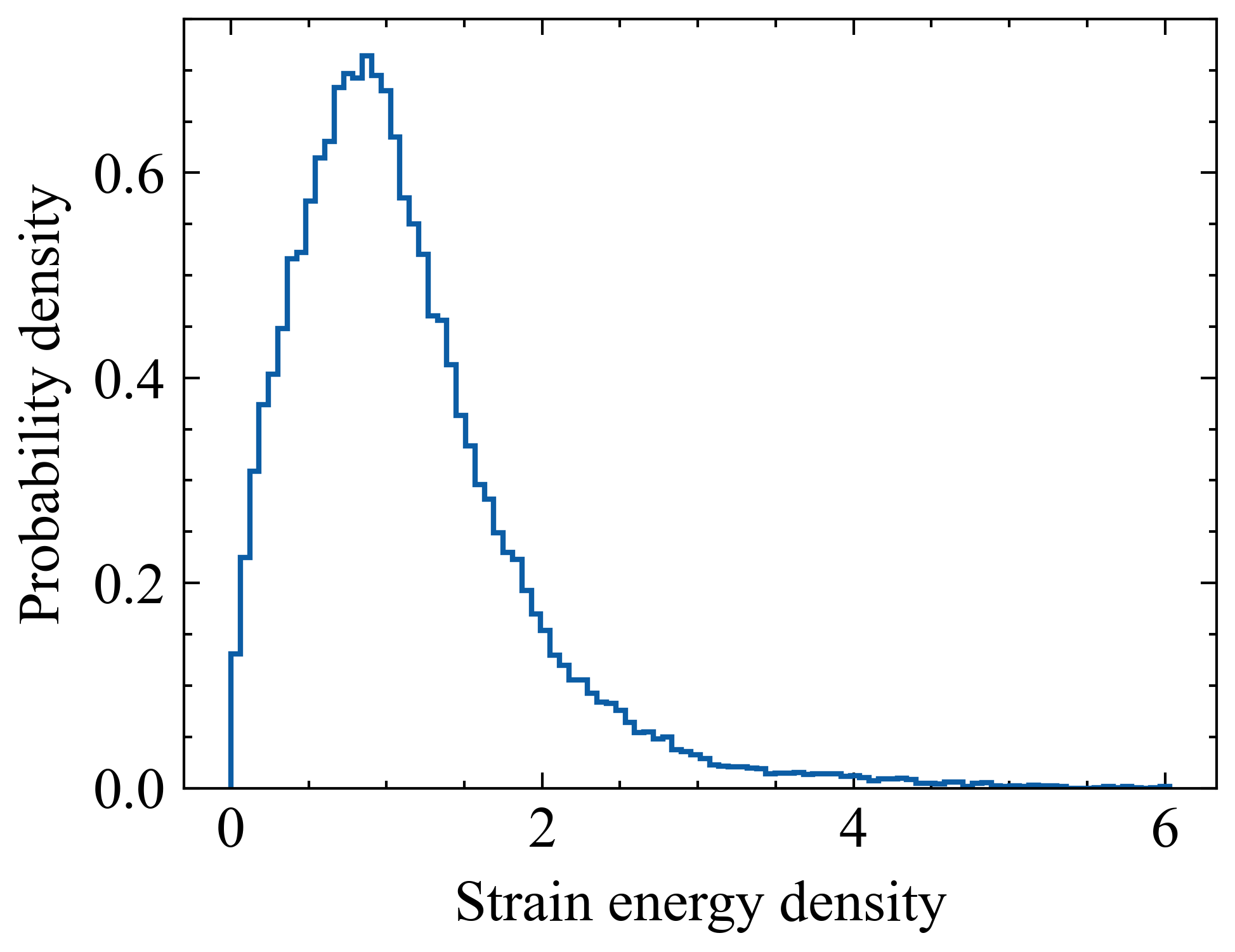}
         \caption{Distribution of strain energy density obtained through MuMFiM simulation of discrete RVE with sampled $F$.}
    \end{subfigure}
    \hfill
    \begin{subfigure}[t]{0.3\textwidth}
        \centering
         \includegraphics[width=\textwidth]{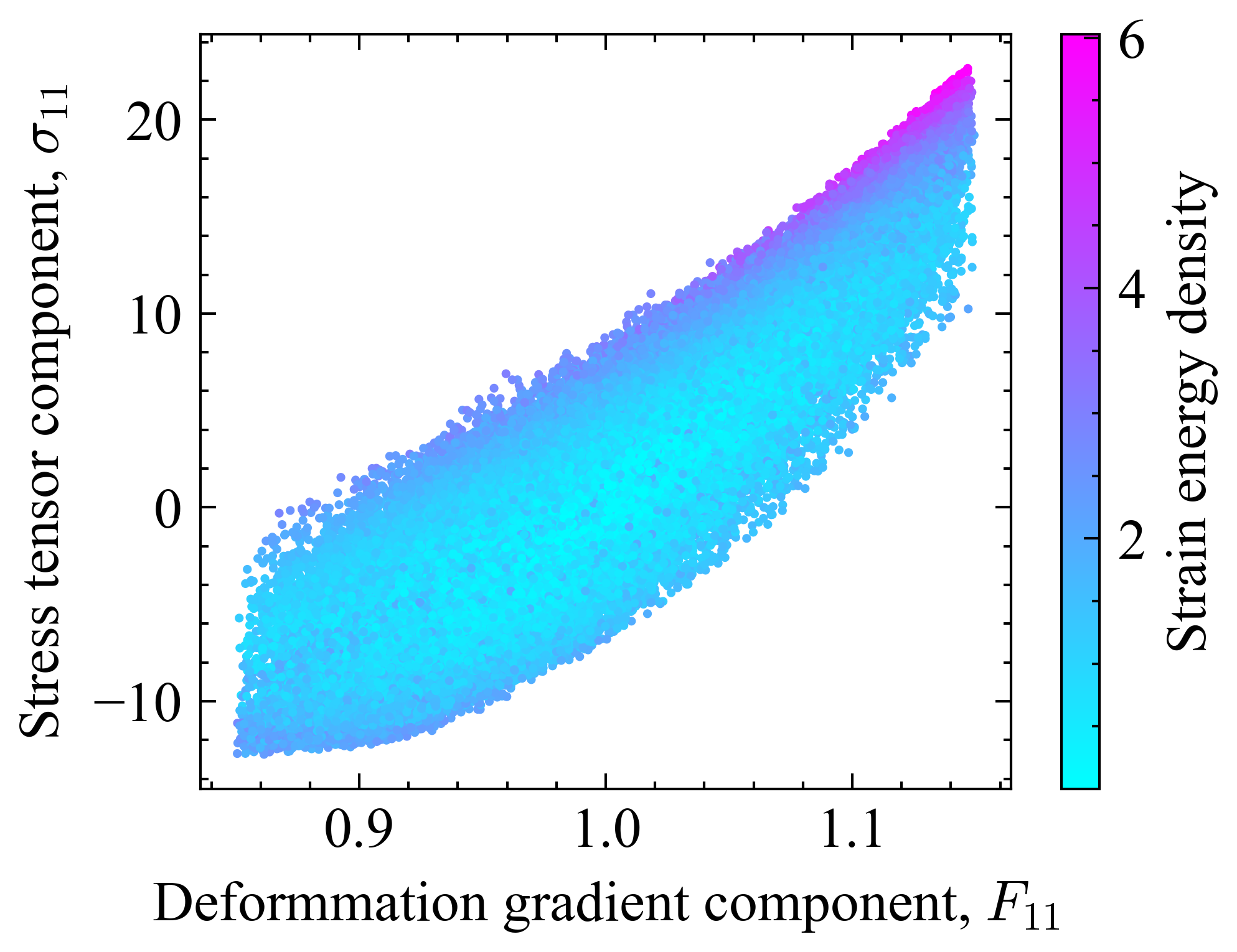}
         \caption{Sample stress vs component $\vb{F}$, showing $\sigma_{11}$ vs $F_{11}$.}
    \end{subfigure}
    \hfill
    \begin{subfigure}[t]{0.3\textwidth}
        \centering
         \includegraphics[width=\textwidth]{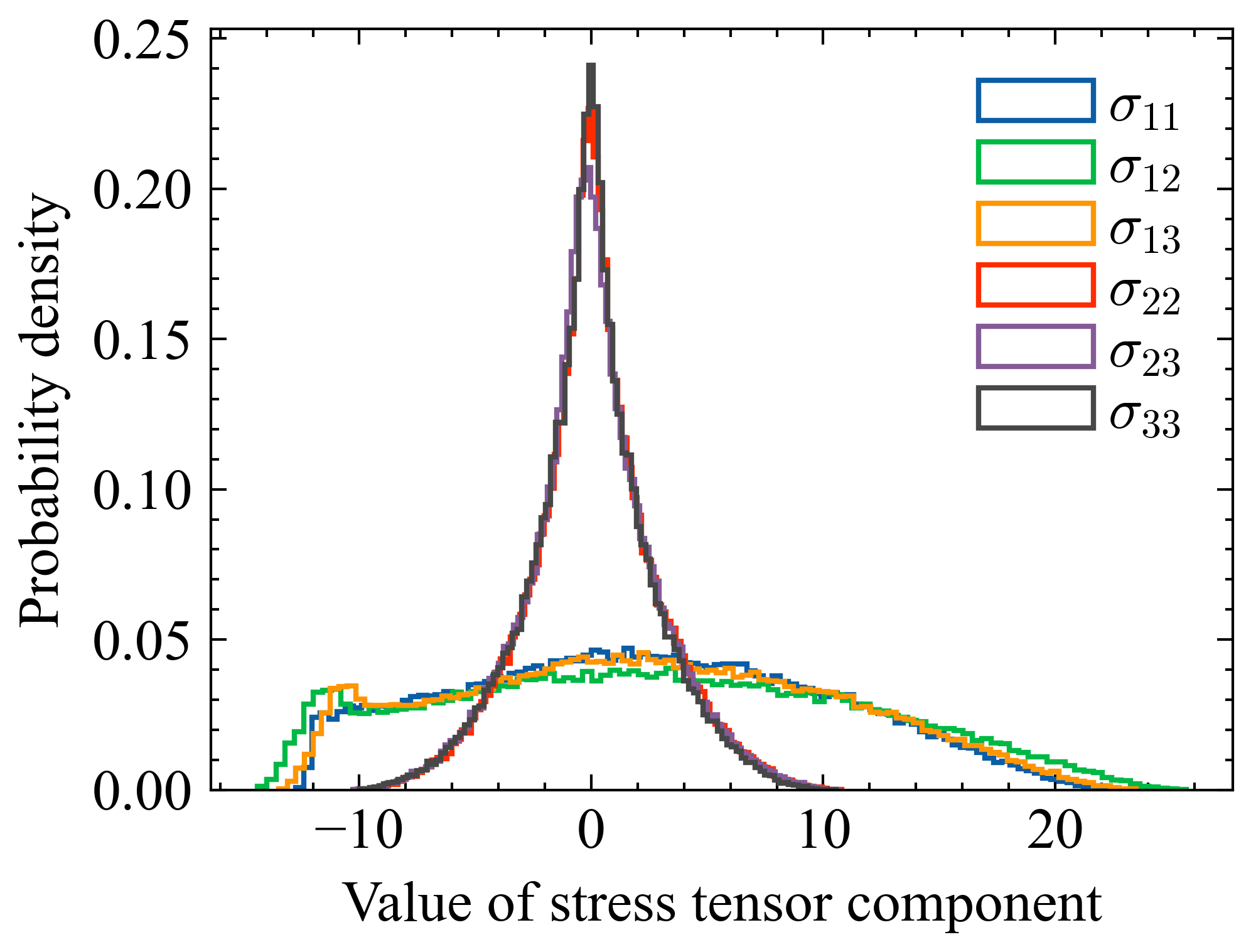}
         \caption{Distribution of components of stress tensor, $\vb{\sigma}$.}
    \end{subfigure}
    
    \caption{Few characteristic results of RVE subjected to sampled $\vb{F}$ through MuMFiM.}
    \label{fig:mumfim_result}
\end{figure}

\FloatBarrier

\section{ICNN Implementation}
Here, we enumerate a few key parameters of the ICNN used in this study. More details are available in the online repository specified in the main paper.

\begin{itemize}
	\item Layer specifications: $k \times 4 \times 4 \times 4 \times 4 \times 4 \times 1$ with pass-through layers for hidden layer neurons. $k$ is the dimension of the input array.
	\item  Number of trainable parameters: 164 when $k = 3$ ($I_1-3$, $I_2-3$, $I_3 - 1$ as input) and 139 when $k = 2$ ($I_1 -3$ and $I_2 -1$ as input)
	\item Activation function: \texttt{Softplus}
    \item Non-negative weights: Enforced through applying \texttt{Softplus} over the weights
	\item Initial learning rate: 0.25
	\item Learning rate schedule: Reduce on the plateau with \texttt{patience = 5}
	\item Batch size: 128
	\item Sample size: 91590 with 80/20 split for training and test
	\item Optimizer: Adam with no weight decay
	\item Other hyperparameters not specified here are set to \texttt{PyTorch v2.3.0}
 \end{itemize}

\section{Error Measures}
Here we provide detailed error measures for various loss functions considered. The error measures we report are $R^2$ and normalized MSE (NMSE). The corresponding values for strain energy density are 0.9953 and 0.0015. They remain approximately the same for different loss functions. The results for various components of stress and stiffness tensor are shown below.

\begin{figure}[h!]
    \centering
    \begin{subfigure}[b]{0.45\textwidth}
        \centering
         \includegraphics[width=\textwidth]{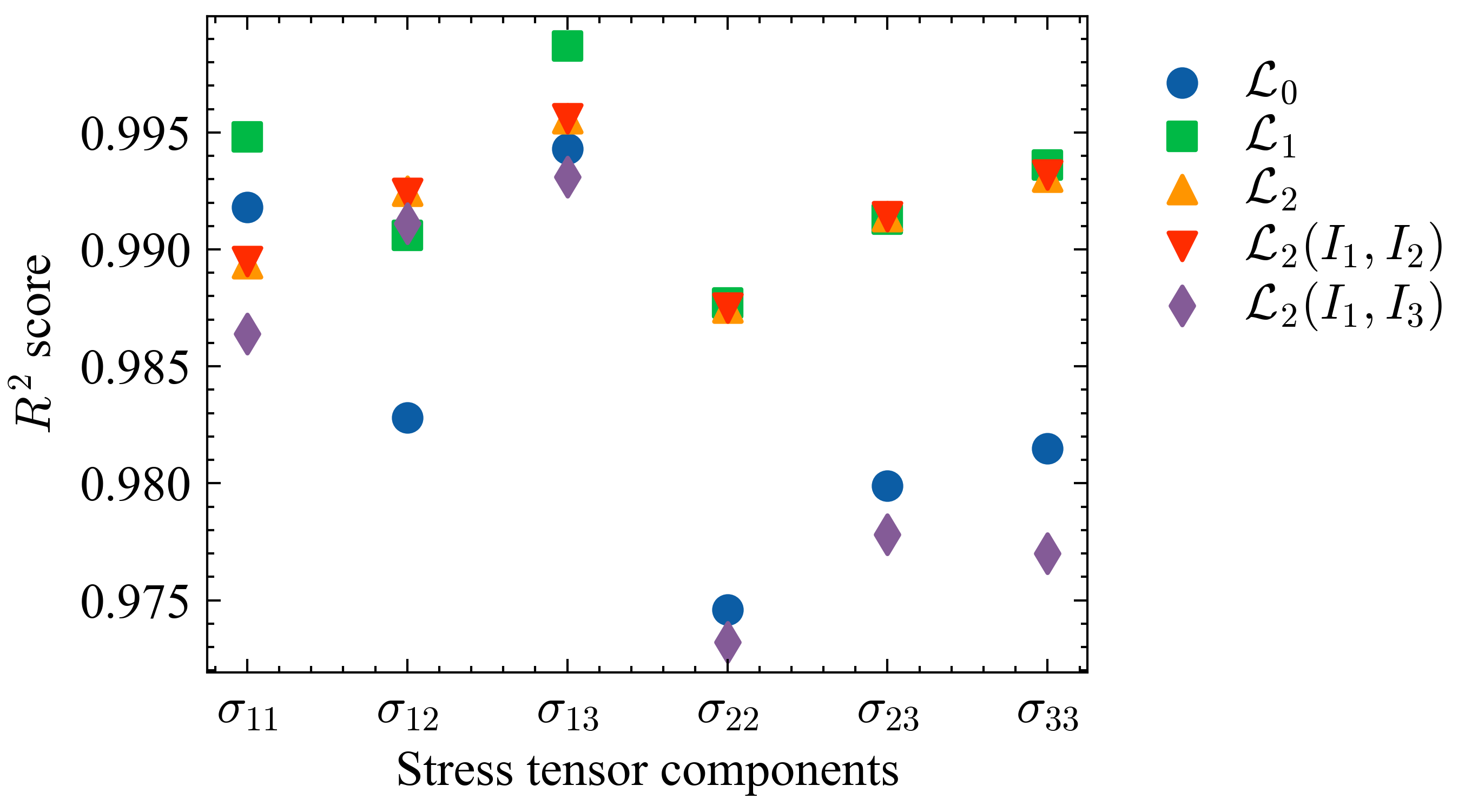}
         \caption{$R^2$ score}
    \end{subfigure}
    \hfill
    \begin{subfigure}[b]{0.45\textwidth}
        \centering
         \includegraphics[width=\textwidth]{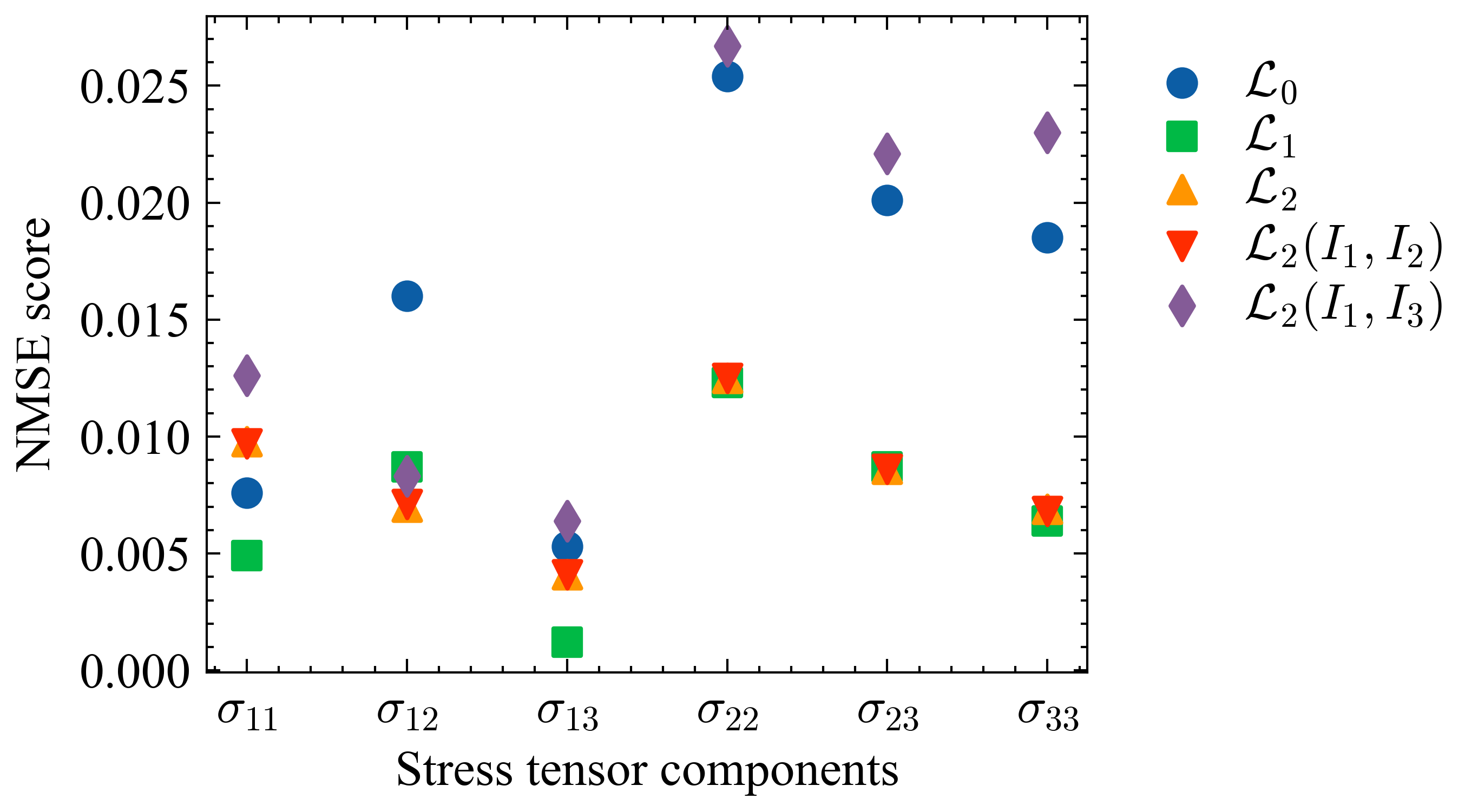}
         \caption{Normalized MSE score}
    \end{subfigure}
    \caption{Averaged error measures of stress components for various loss functions. The input to ICNN is comprised of all isotropic invariants of $\vb{C}$ unless otherwise indicated.}
    \label{fig:err_stress}
\end{figure}

\begin{figure}[h!]
    \centering
    \begin{subfigure}[b]{\textwidth}
        \centering
         \includegraphics[width=\textwidth]{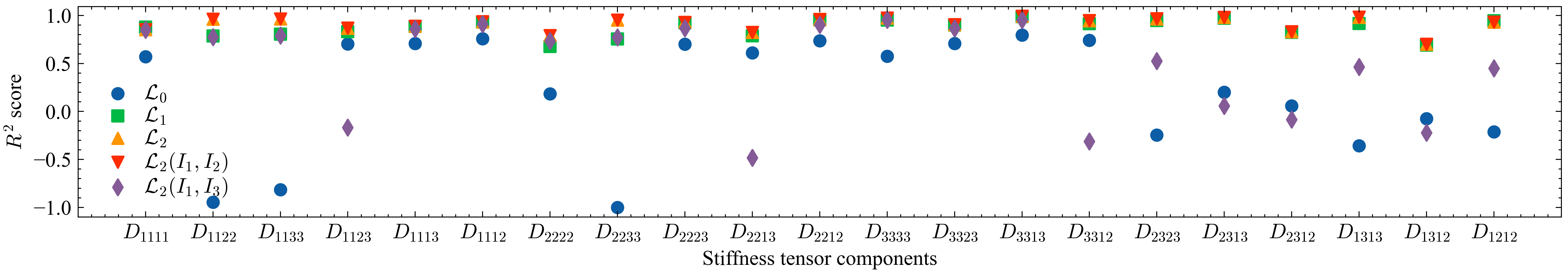}
         \caption{$R^2$ score}
    \end{subfigure}
    \hfill
    \begin{subfigure}[b]{\textwidth}
        \centering
         \includegraphics[width=\textwidth]{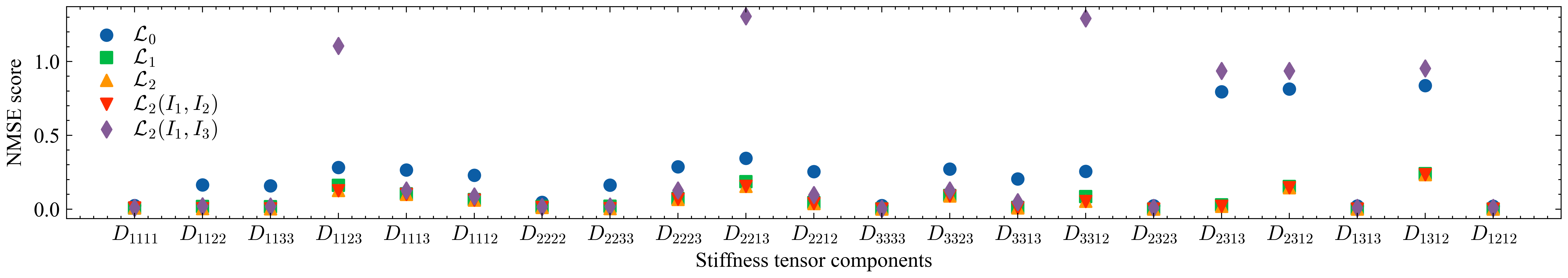}
         \caption{Normalized MSE score}
    \end{subfigure}
    \caption{Averaged error measures of stiffness tensor components for various loss functions. The input to ICNN is comprised of all isotropic invariants of $\vb{C}$ unless otherwise indicated.}
    \label{fig:err_siffness}
\end{figure}